\documentclass{jfm}
\usepackage{graphicx}
\usepackage{epstopdf, epsfig}
\usepackage{url}
\usepackage{dcolumn}
\usepackage{bm}
\usepackage{amsmath}
\usepackage{float}

\shorttitle{Effect of spatial dimension on a model of fluid turbulence}
\shortauthor{D. Clark, R.D.J.G Ho and A. Berera}

\title{Effect of spatial dimension on a model of fluid turbulence}

\author{Daniel Clark\aff{1}
  \corresp{\email{daniel-clark@ed.ac.uk}},
  Richard D. J. G. Ho\aff{2}
 \and Arjun Berera \aff{1}}

\affiliation{\aff{1}School of Physics and Astronomy, University of Edinburgh, JCMB, \\ King’s Buildings, Peter Guthrie Tait Road EH9 3FD, Edinburgh, United Kingdom.
\aff{2}Marian Smoluchowski Institute of Theoretical Physics, Jagiellonian University, Łojasiewicza 11, 30-348, Kraków, Poland}

\begin{document}

\maketitle

\begin{abstract}
 A numerical study of the $d$-dimensional Eddy Damped Quasi-Normal Markovian equations is performed to investigate the dependence on spatial dimension of homogeneous isotropic fluid turbulence. Relationships between structure functions and energy and transfer spectra are derived for the $d$-dimensional case. Additionally, an equation for the $d$-dimensional enstrophy analogue is derived and related to the velocity derivative skewness. Comparisons are made to recent four dimensional direct numerical simulation results. Measured energy spectra show a magnified bottleneck effect which grows with dimension whilst transfer spectra show a varying peak in the non-linear energy transfer as the dimension is increased. These results are consistent with an increased forward energy transfer at higher dimensions, further evidenced by measurements of a larger asymptotic dissipation rate with growing dimension. The enstrophy production term, related to the velocity derivative skewness, is seen to reach a maximum at around five dimensions and may reach zero in the limit of infinite dimensions, raising interesting questions about the nature of turbulence in this limit.
\end{abstract}


\section{Introduction}

Despite more than a century of concentrated effort, fluid turbulence remains 
steadfast as the oldest unsolved problem of classical physics. Much of the 
progress in developing our understanding of turbulence can be traced to 
the work of Kolmogorov and his three 1941 
papers \citep{kolmogorov1941dissipation, kolmogorov1941local, kolmogorov1941degeneration}, in which what has come to be known as the K41 theory was 
first described. The results in these papers are derived for an idealised form of turbulence known as homogeneous and isotropic turbulence (HIT), however they are also remarkably applicable to real world flows under certain conditions. One of the most important predictions of the K41 theory, valid at sufficiently high Reynolds number, is the existence of a range of intermediate sized eddies in the flow referred to as the inertial range, characterised by scale invariance and a constant energy flux. This scale invariance manifests itself clearly in the power law form of the K41 energy spectrum in the inertial range \begin{equation}\label{ek}
E(k) = C \varepsilon^{\frac{2}{3}} k^{-\frac{5}{3}} \;,
\end{equation} where $\varepsilon$ is the constant energy flux, which for stationary turbulence will be equal to the rate of viscous energy dissipation, and $C$ is a universal constant.

The scale invariance of the inertial range is reminiscent of that seen in 
critical phenomena close to the critical point. Following this line of 
argument, there have been numerous analogies comparing turbulence to 
critical phenomena \citep{nelkin1974turbulence, de1975phase, bramwell1998universality, aji2001fluctuations, yakhot2001mean, giuliani2002critical, frisch2012turbulence}. A 
salient feature of many such critical systems is the existence of an 
upper critical dimension, above which fluctuations are suppressed
and the mean field theory values for critical exponents become exact. 
These ideas have their roots in the work of \citet{ginzburg1960vl}, as 
well as that of Wilson and Fisher, in the application of renormalisation group methods 
to critical phenomena \citep{wilson1971renormalization, wilson1972critical}. 
For turbulence, a case can be made that the K41 theory, since it uses the 
mean energy flux in the form of the inertial range energy spectrum, is in 
fact a kind of mean field theory itself \citep{siggia1977origin, bell1978time}.

This interpretation of K41 is interesting in light of the measurement of 
deviations from the exponents predicted by K41 for both the energy spectrum 
and the structure functions. Both intermittency and finite Reynolds number 
effects have been theorised as being responsible for such 
deviations \citep{kolmogorov1962refinement, frisch1978simple, benzi1984multifractal, mccomb2014homogeneous}, which leads to 
equation (\ref{ek}) being re-expressed in the form \begin{equation}
E(k) \propto k^{-\frac{5}{3} - \mu} \;.
\end{equation} From here an analogy can be drawn once more to critical 
phenomena, in which a similar anomalous exponent, which vanishes for mean 
field theory, is seen when looking at two point correlation functions. 
Naturally, this has led to speculation about whether an upper critical 
dimension for turbulence exists and, if so, what its associated value 
would be \citep{rose1978fully, liao1990some, liao1991kolmogorov, nelkin2001does}. Results from a recent 
study \citep{berera2019d} which  performed direct numerical simulation 
(DNS) of four spatial dimensional HIT found, amongst other
results, a suppression of
energy fluctuations in going from three to four dimensions, which has raised further interesting questions 
relating to a critical dimension in turbulence. There have also been 
claims related to, and a small number of studies investigating, 
the possibility of simplification in infinite 
dimensions \citep{kraichnan1974convection, fournier1978infinite, fournier1978d}.

To date only a handful of DNS studies of turbulence in spatial dimensions greater than three have been carried out. The work of \citet{suzuki2005energy} came first with the same group following up on this work in \citep{gotoh2007statistical}. These insightfully motivated studies were focussed on the effect of the spatial dimension on the intermittent nature of turbulence. In both cases due to the computational limitations of the time only decaying turbulence was studied. Here, they found an increased efficiency of the energy transfer in four dimensions compared with three, as well as increased anomalous scaling of the longitudinal structure functions. More recently, in the DNS study of \citet{berera2019d}, which looked at the stationary case at a higher resolution, various measurements performed
pointed to an  increased tendency for energy to be transferred from large to small scales in four dimensions, confirming what had been seen in \citep{gotoh2007statistical}, potentially driven by an increase in vortex stretching.
This interpretation was based on finding a higher velocity derivative skewness. Whether this trend of increased forward energy transfer continues into higher spatial dimensions is an interesting question, as it may indicate the possibility of the turbulent 
dynamics being simplified in higher dimensions. Presently, the computational cost of performing DNS of higher dimensional turbulence is beyond even the largest of supercomputers. Indeed, we are only aware of one study which has investigated beyond four-dimensional turbulence via DNS. In the study by \citet{yamamoto2012local}, a five-dimensional simulation was carried out, though this involved a relatively large lattice spacing, meaning the results are only for very low Reynolds number values where the conditions of K41 are not met. With these  computational considerations in mind, we are forced to turn to closure approximations in order to conduct a feasible study.  

Closure approximations have their roots in quantum field theory (QFT). 
Initially work was pioneered 
by \citet{kraichnan1959structure}, \citet{wyld1961formulation} 
and \citet{edwards1964statistical} in employing QFT methods to
develop a perturbation theory for the Navier-Stokes equations,
and this subsequently led to various approximation schemes.
In this investigation, we will make use of the eddy damped quasi-normal 
Markovian (EDQNM) closure, first described by \citet{orszag1970analytical} 
as a method of achieving realisability in the quasi-normal 
approximation \citep{millionshchikov1941theory}. 
Before proceeding with the details of this method and our calculations,
it is interesting to note that QFT is also a subject in
which the behaviour of systems in different dimensions has
been an area of sustained interest in systems
including string theory, gauge theory, and anti-de Sitter/conformal field theory correspondence (ADS-CFT). 
Hence, aside from the computational tools this area has helped develop for
the field of fluid turbulence, there is also relevance in
appreciating this conceptual point and thus in placing more focus on
understanding the dimensional behaviour of fluid turbulence.

The EDQNM closure scheme has seen 
widespread use in both two and three-dimensional turbulence 
(see \citet{lesieur1987turbulence} for an in-depth review), where it has 
produced numerous qualitative results. The EDQNM 
approximation allows investigation to very high Reynolds number flows 
at relatively low computational cost and has the added benefit 
that extension to any dimension incurs no additional computational expense. 
The EDQNM closure is compatible with the Kolmogorov energy spectrum and 
is well suited for the study of energy transfer in isotropic turbulence. 
Additionally, it was noted by \citet{orszag1974lectures} that the 
quasi-normal approximation is analogous to the random phase approximation 
of many-body physics, 
which is also closely linked to the Gaussian approximation.
Therefore, if higher dimensional turbulence shows
a systematic improvement in agreement 
with the EDQNM approximation, it may in its own right be an
indicator towards a simplification 
in the turbulent dynamics. 

If forward energy transfer does indeed become stronger with higher spatial dimension, this may result in an increased bottleneck effect \citep{falkovich1994bottleneck}. This effect manifests itself as a pile up of energy in the near dissipative range of the flow and has been observed both experimentally \citep{mestayer1982local, saddoughi1994local} and numerically  \citep{kerr1990velocity}. It has been suggested by \citet{herring1982comparative} that this effect is a result of viscosity suppressing the non-linear transfer of energy to the smallest scales. Hence, by varying the spatial dimension of the system it is possible to investigate these claims of viscous energy transfer suppression.

In \citet{berera2019d} a DNS dataset of unprecedented size for four spatial dimensions was developed.  It reached a box size of $512^4$, which in terms of computational demands is similar to $4096^3$, so amongst the larger DNS datasets. Moreover, this was a forced simulation, run for a very long time to achieve good equilibration and adequate time for robust sampling of the data.  The simulation focused on
studying large scale properties of the four dimensional turbulent state. In particular it examined the anomalous dissipation and total energy fluctuation, comparing behaviour between three and four spatial dimensions.  The study found a significant suppression of the energy fluctuation in four dimensions, thus having some qualitative similarities to behaviour found in critical phenomenon. Additionally, the study found an increased velocity derivative skewness and asymptotic dissipation rate in four spatial dimensions compared to three. This is consistent with the interpretation on an enhanced forward energy transfer in four dimensions when compared to three dimensions. The study by \citet{berera2019d} also served the purpose of providing a baseline of results that will be useful for comparison in any future DNS studies in four spatial dimensions.  For all these reasons it is important to have an independent check of the very new type of results seen in that study. Where possible this paper will examine these quantities in four spatial dimensions using the EDQNM approximation independently confirming several of the DNS results in \citet{berera2019d}. We will then go further and examine similar results in higher dimensions. The structure of this paper is  as follows: in Section \ref{theory} we outline a number derivations  for $d$-dimensional turbulence in the Navier-Stokes equations.  Section \ref{methods} introduces the EDQNM closure model used for this work. Section \ref{results} presents the results of our numerical study and finally, section \ref{conclusion} discusses the interpretation and possible implications of these results.

\section{Theory}\label{theory}

The incompressible Navier-Stokes equations (NSE) can be expressed for spatial dimension  $d \geq 2$ as \begin{equation}\label{nsereal}
\begin{split}
\partial_t \bm{u} + \bm{u} \cdot \bm{\nabla} \bm{u} &= -\bm{\nabla} P + \nu \nabla^2 \bm{u}\;, \\
\bm{\nabla} \cdot \bm{u} &= 0\;.
\end{split}
\end{equation} In the above $\bm{u}(\bm{x},t)$ is the velocity 
field, $P(\bm{x},t)$ is the pressure field, $\nu$ is the kinematic 
viscosity and $\bm{\nabla} \cdot \bm{u} = 0$ is the incompressibility 
condition, which allows us to set the fluid density to unity.

Here we discuss the relevant quantities and definitions that
are utilised throughout the rest of the paper.
For brevity 
in the following we will drop the explicit time dependence. For our 
purposes, we are primarily interested in the second and third order two 
point velocity correlations, which are given by \begin{equation}
\begin{split}
C_{\alpha \beta}(\bm{r}) &= \langle u_{\alpha}(\bm{x}) u_{\beta}(\bm{x+r})\rangle\;, \\
C_{\alpha \beta \gamma}(\bm{r}) &= \langle u_{\alpha}(\bm{x}) u_{\beta}(\bm{x}) u_{\gamma}(\bm{x+r})\rangle \;,
\end{split}
\end{equation} with $\alpha, \beta, \gamma = 1,\dots,d$. More specifically, we will be focussed on the second and third order longitudinal correlations defined as \begin{equation}
\begin{split}\label{corlong}
C_{LL} &= \frac{r_{\alpha} r_{\beta}}{r^2}C_{\alpha \beta}(\bm{r}) = u^2 f(r), \\
C_{LL,L} &= \frac{r_{\alpha} r_{\beta} r_{\gamma}}{r^3}C_{\alpha \beta \gamma}(\bm{r}) = u^3 K(r)\;,
\end{split}
\end{equation} where $f(r)$ and $K(r)$ are scalar correlation functions and $u$ is the RMS velocity. These functions are intimately related to the longitudinal structure functions of the same order. It will prove useful to introduce the Fourier transform of equation (\ref{nsereal}) \begin{equation}
\left(\partial_t + \nu k^2\right) u_{\alpha}(\bm{k}) = \frac{1}{2i}P_{\alpha \beta \gamma}(\bm{k}) \int \mathrm{d}\bm{p} \, u_{\beta}(\bm{p})u_{\gamma}(\bm{k-p})\;,
\end{equation} where $P_{\alpha \beta \gamma}(\bm{k}) = k_{\beta} P_{\alpha \gamma}(\bm{k}) + k_{\gamma}P_{\alpha \beta}(\bm{k})$ is the inertial transfer operator and $P_{\alpha \beta}(\bm{k}) = \delta_{\alpha \beta} - k_{\alpha} k_{\beta}/k^2$ is the projection operator which imposes the incompressibility condition. Homogeneity requires that the corresponding second order velocity correlation in Fourier space takes the form \begin{equation}
C_{\alpha \beta}(\bm{k}) = \langle u_{\alpha}(\bm{k})u_{\beta}(\bm{-k}) \rangle \;.
\end{equation}  It can be shown that this correlator is related to the energy spectrum, $E(k)$, through \begin{equation}
\langle u_{\alpha}(\bm{k})u_{\beta}(\bm{-k}) \rangle = \frac{2P_{\alpha \beta}(\bm{k})E(k)}{(d-1)A_{d}k^{d-1}} \;,
\end{equation} where $A_{d} = 2\left( (\pi)^{d/2}/\Gamma(d/2)\right)$ is the surface area of a $d$-dimensional unit sphere. We can then form an equation for $E(k)$ \begin{equation}\label{nsef}
\begin{split}
\left(\partial_t + 2\nu k^2\right)E(k) &=  \frac{i A_{d} k^{d-1}}{2}  P_{\alpha \beta \gamma}(\bm{k}) \mathcal{A}_{\alpha \beta \gamma}(\bm{k}) \\ &= T(k) \;.
\end{split}
\end{equation} Here, $T(k)$ is the energy transfer spectrum and $\mathcal{A}_{\alpha \beta \gamma}(\bm{k})$ is defined as \begin{equation}\label{A}
\mathcal{A}_{\alpha \beta \gamma}(\bm{k}) = \int \mathrm{d}\bm{p} \, C_{\alpha \beta \gamma}(\bm{k},\bm{p},-\bm{k}-\bm{p}) \;,
\end{equation} where $C_{\alpha \beta \gamma}(\bm{k},\bm{p},-\bm{k} -\bm{p}) = \langle u_{\alpha}(\bm{k})u_{\beta}(\bm{p}) u_{\gamma}(-\bm{k}-\bm{p})\rangle$ 
is the spectral third order velocity field moment.  More detailed
derivations and interpretation of these quantities 
can be found in \citep{mccomb1990physics}. 

\subsection{Second Order Structure Function}

The dimensional dependence of HIT is well elucidated in the form of the velocity correlation functions, and thus in turn the structure functions. The longitudinal structure function of order $n$ is given by \begin{equation}
S_{n}^{(d)} = \left\langle \bm{u}(\bm{x}+\bm{r}) - \bm{u}(\bm{x}) \right\rangle \cdot \frac{\bm{r}}{r}\;.
\end{equation} Now, we derive the relationship between the second and third order longitudinal structure functions and the energy and transfer spectra respectively. As a result, we are then able to evalutate these structure functions in our numerical EDQNM results. The method used here is the $d$-dimensional extension of that used in \citet{bos2012reynolds}. We begin by considering the second order longitudinal structure function which can be expressed as \begin{equation}\label{struct2}
S_{2}^{(d)}(r) = 2\left(u^2 - C_{LL}\right) = 2\left( u^2 - \frac{r_{\alpha} r_{\beta}}{r^2}C_{\alpha \beta}(\bm{r}) \right)\;.
\end{equation} These correlations are related to their spectral analogues via a $d$-dimensional inverse Fourier transform \begin{equation}
C_{\alpha \beta}(\bm{r}) = \int \mathrm{d}\bm{k} \, C_{\alpha \beta}(\bm{k}) e^{i\bm{k}\cdot \bm{r}} = \int \mathrm{d}\bm{k} \, \frac{2P_{\alpha \beta}(\bm{k})E(k)}{(d-1)A_{d}k^{d-1}} e^{i\bm{k}\cdot \bm{r}}\;. 
\end{equation} In HIT, these transforms are simplified by the fact that the correlators must be spherically symmetric, thus we have \begin{equation}\label{cll}
\begin{split}
C_{LL} &= \int \mathrm{d}\bm{k} \, \frac{2(1-\cos^2 \theta)E(k)}{(d-1)S_{d}k^{d-1}} e^{i\bm{k}\cdot \bm{r}}\\  &= \int_{0}^{\infty} \mathrm{d}k \, \frac{2E(k)}{(d-1)A_{d}}  A_{d-1}\int_{0}^{\pi}\mathrm{d}\theta \, e^{ikr\cos\theta}\sin^{d-2}\theta(1-\cos^2 \theta) \\ &= 2^{\frac{d}{2}}\Gamma\left(\frac{d}{2}\right)\int_{0}^{\infty} \mathrm{d}k \, E(k)\frac{J_{\frac{d}{2}}(kr)}{(kr)^{\frac{d}{2}}}\;.
\end{split}
\end{equation} Where $\theta$ is defined as the angle between $\bm{k}$ and $\bm{r}$, such that $\bm{k} \cdot \bm{r} = kr\cos \theta$ and hence $r_\alpha r_\beta P_{\alpha \beta}(\bm{k}) = r^2 \left(1-\cos^2 \theta\right)$, and $J_{n}(x)$ is the $n$-th order Bessel function of the first kind. Additionally, for $u^2$ we find \begin{equation}
u^2  = \frac{2}{d}\int_{0}^{\infty}\mathrm{d}k \, E(k)\;.
\end{equation} Inserting these two results into equation (\ref{struct2}) yields \begin{equation}\label{2ndord}
S_{2}^{(d)}(r) = 2\int_{0}^{\infty} \mathrm{d}k \, E(k) \left[\frac{2}{d} -2^{\frac{d}{2}}\Gamma\left(\frac{d}{2}\right) \frac{J_{\frac{d}{2}}(kr)}{(kr)^{\frac{d}{2}}}\right]\;,
\end{equation} thus we have related $S_{2}^{(d)}(r)$ to $E(k)$. It can be verified that this gives the standard results for two and three dimensions as found in \citet{davidson2015turbulence}. Additionally, this result allows us to determine the integral length scale in $d$-dimensional HIT. The integral length scale is defined \citep{batchelor1953theory} as \begin{equation}
L_d = \int_{0}^{\infty} \mathrm{d}r f(r)\;,
\end{equation} where $f(r)$ is the scalar correlation function defined in equation (\ref{corlong}), such that $f(r) = C_{LL}/u^2$. Therefore, for the $d$-dimensional case we have \begin{equation}\label{L}
\begin{split}
L_{d} = \frac{2^{\frac{d}{2}}\Gamma\left(\frac{d}{2}\right)}{u^2} \int_{0}^{\infty} \mathrm{d}k \, E(k) \int_{0}^{\infty} \mathrm{d}r \, \frac{J_{\frac{d}{2}}(kr)}{(kr)^{\frac{d}{2}}} = \frac{\Gamma\left(\frac{d}{2}\right)\sqrt{\pi}}{\Gamma\left(\frac{d+1}{2}\right)u^2}\int_{0}^{\infty}\mathrm{d}k \, E(k)k^{-1}\;.
\end{split}
\end{equation} We will find this expression useful when defining our integral scale Reynolds number. It is also possible to generalise the Taylor mircroscale \citep{taylor1935statistical}, $\lambda_d$, which gives the average size of the dissipative eddies, to $d$-dimensions. This length scale is defined through fitting a parabola to the small $r$ expansion of the scalar longitudinal correlation function $f(r)$, i.e. \begin{equation}
f(r) = 1 - \frac{r^2}{2\lambda_{d}^2} + \mathcal{O}(r^4).
\end{equation} From equations (\ref{corlong}) and (\ref{cll}) we find through expansion for small $r$ \begin{equation}\label{smallr2nd}
f(r) = \frac{C_{LL}}{u^2} = 1 - \frac{r^2}{d(d+2)u^2}\int_{0}^{\infty} \mathrm{d}k \, k^2 E(k) + \mathcal{O}(r^4)\;,
\end{equation} where if we recall that \begin{equation}
\varepsilon = 2 \nu \int_{0}^{\infty} \mathrm{d}k \, k^2 E(k)\;,
\end{equation} we then have \begin{equation}
f(r) = 1 - \frac{\varepsilon r^2}{2d(d+2)\nu u^2} + \mathcal{O}(r^4)\;.
\end{equation} Hence, the Taylor microscale in $d$-dimensions is given by \begin{equation}
\lambda_{d} = \sqrt{\frac{d(d+2) \nu }{\varepsilon}}u\;.
\end{equation} 

\subsection{Third Order Structure Function}

A similar analysis  can be performed for the third order structure function, $S_{3}(r)$, whereby it is related to the transfer spectrum. We begin with the relation \begin{equation}
\begin{split}
S_{3}^{(d)}(r) = 6C_{LL,L} = 6\frac{r_{\alpha} r_{\beta} r_{\gamma}}{r^3}C_{\alpha \beta \gamma}(\bm{r})\;, 
\end{split}
\end{equation} and find that upon Fourier transform we have \begin{equation}
\mathcal{F}\left[ C_{\alpha \beta \gamma}(\bm{r}) \right] = \mathcal{A}_{\alpha \beta \gamma}(\bm{k})\;,
\end{equation} with $A_{\alpha \beta \gamma}(\bm{k})$ defined as in equation (\ref{A}). From here, we can observe that \begin{equation}
S_{3}^{(d)}(r) = 6\frac{r_{\alpha} r_{\beta} r_{\gamma}}{r^3} \int \mathrm{d}\bm{k} \, \mathcal{A}_{\alpha \beta \gamma}(\bm{k}) e^{i\bm{k}\cdot\bm{r}}\;.
\end{equation}  Now all that remains is to express this in terms of the transfer function and perform the Fourier integrals. From equation (\ref{nsef}) we have \begin{equation}
\frac{iA_{d} k^{d-1}}{2} P_{\alpha \beta \gamma}(\bm{k})\mathcal{A}_{\alpha \beta \gamma}(\bm{k}) = T(k)\;,
\end{equation} and, as $\mathcal{A}_{\alpha \beta \gamma}(\bm{k})$ is a third rank solenoidal tensor symmetric in the indices $\beta$ and $\gamma$, we must have $\mathcal{A}_{\alpha \beta \gamma}(\bm{k}) = P_{\alpha \beta \gamma}(\bm{k})\mathcal{A}(k) $.  Thus by evaluating the product $P_{\alpha \beta \gamma}(\bm{k})P_{\alpha \beta \gamma}(\bm{k}) = 2(d-1)k^2$ we have \begin{equation}
\mathcal{A}_{\alpha \beta \gamma}(\bm{k}) = \frac{P_{\alpha \beta \gamma}(\bm{k}) T(k)}{i(d-1)A_{d}k^{d+1}}\;.
\end{equation} Hence, for $S_{3}^{(d)}(r)$ we have the following \begin{equation}\label{3rdord}
S_{3}^{(d)}(r) = 6\frac{r_{\alpha} r_{\beta} r_{\gamma}}{r^3} \int \mathrm{d}\bm{k} \frac{P_{\alpha \beta \gamma}(\bm{k}) T(k)}{i(d-1)A_{d}k^{d+1}} e^{i\bm{k}\cdot\bm{r}}\;.
\end{equation} 
In the same way as we did for $S^{(d)}_2(r)$ we can evaluate this integral by taking $\theta$ to be the angle between $\bm{k}$ and $\bm{r}$, which upon doing so we find \begin{equation}
\frac{r_{\alpha} r_{\beta} r_{\gamma}}{r^3}P_{\alpha \beta \gamma}(\bm{k}) = 2k\cos \theta \left(1-\cos^2 \theta\right)\;.
\end{equation} We then once more evaluate all but two of the $d$ Fourier integrals to obtain \begin{equation}
S_{3}^{(d)}(r) = \frac{12 A_{d-1}}{i(d-1)A_d}\int_{0}^{\infty} \mathrm{d}k \, \frac{T(k)}{k}\int_{0}^{\pi} \mathrm{d}\theta \, \cos \theta \left(1-\cos^2 \theta\right)\sin^{d-2}\theta e^{ikr\cos \theta}\;.
\end{equation} The inner integral is formidable, however it can be evaluated using computer algebra software. The result when restricted to integer dimensions is then found to be \begin{equation}\label{3struct}
S_{3}^{(d)}(r) = 3\Gamma\left(\frac{d}{2}\right)r\int_{0}^{\infty} \mathrm{d}k \, 2^{1+\frac{d}{2}}T(k) \frac{J_{1+\frac{d}{2}}(kr)}{\left(kr\right)^{1+\frac{d}{2}}}\;.
\end{equation} 

As a check we compare this result with the case for $d=2$ derived in \citep{cerbus2017third} \begin{equation}
S_{3}^{(2)}(r) =  12 r\int_{0}^{\infty} \mathrm{d}k \, T(k) \frac{J_2 (kr)}{\left(kr\right)^2}\;.
\end{equation} Now, clearly upon inserting $d=2$ into equation (\ref{3struct}) we recover the result above. Furthermore using properties of Bessel functions it can also easily be shown that for $d=3$ the above reduces to the expected expression as seen in \citep{bos2012reynolds}. 

Finally, we consider a small $r$ expansion of the third order structure function in $d$ dimensions \begin{equation}
S_{3}^{(d)} = \frac{12r}{d(d+2)}\int_{0}^{\infty}\mathrm{d}k \, T(k) - \frac{6r^3}{d(d+2)(d+4)} \int_{0}^{\infty}\mathrm{d}k \, k^2 T(k) + \mathcal{O}(r^5)\;.
\end{equation} From this expansion and the conservation properties of $T(k)$, we can see $S_{3}^{(d)}(r) \sim r^3$ for small $r$ and $d\geq 3$. In two dimensions the second term also vanishes as a result of enstrophy conservation. This expansion is also of practical use for evaluation of $S_{3}^{(d)}(r)$ for very small $r$ numerically, where floating point arithmetic errors can arise

\subsection{Enstrophy Production and Skewness}\label{sectionskew}

Vorticity and enstrophy play an important role in the behaviour of two and three dimensional turbulence. Enstrophy production is also known to be linked to the velocity derivative skewness, hereafter refered to simply as the skewness, of the flow. To generalise these concepts to arbitrary spatial dimension, we first introduce the vorticity 2-form \begin{equation}
\Omega_{\alpha \beta}(\bm{x}) = \partial_{\alpha} u_{\beta}(\bm{x}) - \partial_{\beta} u_{\alpha}(\bm{x})\;.
\end{equation} We can then re-express equation (\ref{nsereal}) using $\Omega_{\alpha \beta}$  as \begin{equation}\label{ndnse}
\partial_t u_{\alpha} + \Omega_{\beta \alpha} u_{\beta} = -\partial_{\alpha} \left(P + \frac{u^2}{2}\right) + \nu \nabla^2 u_{\alpha}\;,
\end{equation} which is valid in any dimension, and is equivalent to the rotational form of the NSE in three dimensions. Using equation (\ref{ndnse}) we are then able to derive an evolution equation for $\Omega_{\alpha \beta}(\bm{x})$ \begin{equation}\label{vort}
\partial_t \Omega_{\alpha \beta} + u_{\gamma} \partial_{\gamma} \Omega_{\alpha \beta} + \Omega_{\alpha \gamma}S_{\gamma \beta} + \Omega_{\gamma \beta}S_{\alpha \gamma} = \nu \nabla^2 \Omega_{\alpha \beta}\;,
\end{equation} where $S_{\alpha \beta}(\bm{x}) = \left(\partial_{\alpha} u_{\beta}(\bm{x}) + \partial_{\beta} u_{\alpha}(\bm{x})\right)/2$ is the strain tensor. Enstrophy in three dimensions is defined in terms of the vorticity, $\bm{\omega}(\bm{x})$, as \begin{equation}
Z(t) = \frac{1}{2}\langle \omega_{\alpha} \omega_{\alpha} \rangle = \int_{0}^{\infty} \mathrm{d}k \, k^2 E(k) \;,
\end{equation} 
where for this case we also have 
$\omega_{\alpha} = \epsilon_{\alpha \beta \gamma}\Omega_{\beta \gamma}/2$,
which suggests the correct form of enstrophy in terms of the 
two form is \begin{equation}\label{enstro}
Z(t) = \frac{1}{4}\langle \Omega_{\alpha \beta}^2 \rangle = \int_{0}^{\infty}\mathrm{d}k \, k^2 E(k)\;.
\end{equation} To be confident this second equality holds, we will first form an equation for the evolution of $u^2$ using the vorticity 2-form \begin{equation}
\frac{1}{2}\partial_t u_{\alpha}u_{\alpha} + u_{\alpha}\Omega_{\beta \alpha} u_{\beta} = -\partial_{\alpha}\left(P + \frac{u^2}{2}\right)u_{\alpha} + \nu \partial_{\beta}u_{\alpha}\Omega_{\beta \alpha} - \frac{\nu}{2}\Omega_{\alpha \beta}^{2}\;.
\end{equation} Upon averaging and invoking homogeneity we find \begin{equation}\label{energyeq}
\partial_t E(t) = -\frac{\nu}{2}\langle \Omega_{\alpha \beta}^{2}\rangle\;,
\end{equation} which, when compared to the standard result, is \begin{equation}
\langle \Omega_{\alpha \beta}^{2}\rangle = 2\langle\left(\partial_{\beta}u_{\alpha}\right)^2 \rangle\;.
\end{equation} From here, it can be shown that indeed the second equality in equation (\ref{enstro}) holds as \begin{equation}
\langle\left(\partial_{\beta}u_{\alpha}\right)^2 \rangle = -\langle u_{\alpha}\nabla^2 u_{\alpha} \rangle = 2\int_{0}^{\infty}\mathrm{d}k \, k^2 E(k)\;,
\end{equation} and the equality is proved. Thus, we are confident equation (\ref{enstro}) is a consistent generalisation of enstrophy to all dimensions. 

In order to relate the production of enstrophy to skewness, we require an equation for our generalised enstrophy, which we can obtain from the $d$-dimensional vorticity equation above \begin{equation}\label{ensteq}
\partial_t Z(t) = -\frac{1}{2}\langle \Omega_{ij}\Omega_{ik}S_{kj} + \Omega_{ij}\Omega_{kj}S_{ik} \rangle - \frac{\nu}{2}\langle\Omega_{ij}\nabla^2 \Omega_{ij}\rangle\;.
\end{equation} Following steps similar to those for the enstrophy, we can show that the palinstrophy generalises as \begin{equation}
P(t) = \frac{1}{4}\langle\Omega_{ij}\nabla^2 \Omega_{ij}\rangle = \int_{0}^{\infty}\mathrm{d}k \, k^4 E(k)\;.
\end{equation} Now, in order to express equation (\ref{ensteq}) in terms of the skewness, we consider the von K\'{a}rm\'{a}n-Howarth equation \citep{de1938statistical} in $d$-dimensions expressed in terms of the second and third order two point longitudinal correlations \begin{equation}\label{khe}
\partial_t C_{LL} = \frac{1}{r^{d+1}}\partial_r \left[r^{d+1}C_{LL,L} \right] + \frac{2\nu}{r^{d+1}}\partial_r\left[r^{d+1}\partial_r C_{LL}\right]\;.
\end{equation} From the preceeding discussion we now recognise the integrals in equation (\ref{smallr2nd})  as being the total energy, enstrophy and palinstrophy, hence we have \begin{equation}
C_{LL} = \frac{2}{d}E(t) - \frac{r^2}{d(d+2)}Z(t) + \frac{r^4}{4d(d+2)(d+4)}P(t) + \mathcal{O}(r^{5})\;.
\end{equation} Using this expansion in equation (\ref{khe}), and the fact that $C_{LL,L} \sim r^3$ for small $r$, we find to zeroth order in $r$ \begin{equation}
\partial_t E(t) = - 2\nu Z(t)\;.
\end{equation} This is entirely equivalent to equation (\ref{energyeq}) and represents the decay of energy in turbulent flows without external forcing. Continuing now to $\mathcal{O}(r^2)$ \begin{equation}
\partial_t Z(t) = -\left. d(d+2)(d+4)\frac{C_{LL,L}}{r} \right|_{r\rightarrow 0} -2 \nu P(t)\;.
\end{equation} Both these expressions are consistent with what is derived directly from the NSE (see \citet{davidson2015turbulence} for the two and three dimensional cases). Also, we note here that, since this derivation required $C_{LL,L}\sim r^3$ for small $r$, in two dimensions the first term on the right hand side vanishes. Recalling that $S_{3}^{(d)}(r) = 6C_{LL,L}$ and the skewness, $S_0$, can be expressed as \begin{equation} \label{skew}
S_0 =  \left. \frac{S_{3}^{(d)}(r)}{\left[S_{2}^{(d)}(r)\right]^{\frac{3}{2}}}\right|_{r\rightarrow 0} \;,
\end{equation} we then write the enstrophy equation as \begin{equation}\label{enstroskew}
\partial_t Z(t) = -S_{0}\Lambda(d)Z(t)^{\frac{3}{2}} -2 \nu P(t)\;,
\end{equation} where the $\mathcal{O}(r^2)$ term of the small $r$ expansion of $S_{2}^{(d)}(r)$ has been used and the dimensional factor is \begin{equation}
\Lambda(d) = \frac{(d+4)}{3}\sqrt{\frac{2}{d(d+2)}}\;.
\end{equation}  The dimensional factor, $\Lambda(d)$, in this equation is a decreasing function of $d$ with an asymptotic limit of $\sqrt{2}/3$. As such, with increasing dimension a larger skewness is required to generate the same amount of enstrophy. 

In the above, we have demonstrated that, in any dimension, enstrophy production is governed by the action of the strain field on the generalised vorticity. This action can be thought of as the stretching and folding of structures analogous to vortices in all dimensions. From another viewpoint, this stretching is seen to be caused by a non-zero skewness. However, a larger skewness value is required with increasing dimension to produce the same level of vortex stretching.

We can also consider the skewness as a function of dimension using equation (\ref{skew}) and the small $r$ expansions for the structure functions. This gives \begin{equation}\label{dskewness}
S_{0}(d) = -\frac{1}{\Lambda(d)}\int_{0}^{\infty}\mathrm{d}k \, k^2 T(k) \left[\int_{0}^{\infty}\mathrm{d}k \, k^2 E(k)\right]^{-\frac{3}{2}} \;.
\end{equation} 
This result will be used in section \ref{results}
in order to measure the dimensional dependence of $S_0$.

\section{Methods}\label{methods}

As was highlighted earlier in this work, the cost of DNS in dimensions higher than four is prohibitive. Therefore, in order to study the effects of spatial dimension on turbulent fluid flows, we utilise the $d$-dimensional EDQNM closure approximation. Under this approximation we find the equation for the time evolution of the energy spectrum is \begin{equation}\label{edqnmd}
\begin{split}
\left(\partial_t + 2\nu k^2\right) E(k) = &8K_d \iint\limits_{\Delta k} \mathrm{d}p \, \mathrm{d}q \, \frac{k}{pq}b^{(d)}_{kpq}\theta_{kpq}(t) \\ &\times \bigg[\sin^{d-3}(\alpha) k^2 E(p)E(q) - \sin^{d-3}(\beta) p^2 E(q)E(k)\bigg] + f(k)\;.
\end{split}
\end{equation} It should be noted that the $kpq$ subscripts are simply labels and not indices, so no summation is implied. For a derivation of equation (\ref{edqnmd}), see \citet{lesieur1987turbulence} or \citet{sagaut2008homogeneous} for the three dimensional case, and \citet{rose1978fully} for the extension to arbitrary dimension. In equation (\ref{edqnmd}) the integration is performed over wave-vector triads which can form triangles, i.e. $\lbrace \bm{k}, \bm{p}, \bm{q} \rbrace$ satisfying  $\bm{k}+\bm{p}+\bm{q}=\bm{0}$, and we have \begin{equation}
b^{(d)}_{kpq} = \frac{p}{2k}\left((d-3)Z + (d-1)XY + 2Z^3 \right) \;,
\end{equation} where $X, Y$ and $Z$ are the cosines of the angles $\alpha, \beta$ and $\gamma$ which lie opposite the sides $k,p$ and $q$ respectively. As such, $b^{(d)}_{kpq}$, coupled with the two sine terms, contains all the information regarding the geometry of the triadic interactions in the EDQNM closure. We also have the triad relaxation time \begin{equation}
\theta_{kpq}(t) = \frac{1 - e^{-\left(\mu_{k} +\mu_{p} + \mu_{q} + \nu(k^2 + p^2 + q^2)\right)t}}{\mu_{k} + \mu_{p} + \mu_{q} + \nu(k^2 + p^2 + q^2)}\;,
\end{equation} in which the eddy damping rate is given by \begin{equation}\label{eddamp}
\mu_{k} =  \lambda_1 \sqrt{\int_{0}^{k} \mathrm{d}s \, s^2 E(s,t)}\;. 
\end{equation} Finally the dimensional factor $K_d$ is given by \begin{equation}
K_d = \frac{A_{d-1}}{(d-1)^2 A_d}\;.
\end{equation} In these equations there exists a free parameter $\lambda_1$, which can be shown to set the Kolmogorov constant \citep{andre1977influence}, see Appendix \ref{appA} for further details. For certain applications setting the value of the Kolmogorov constant will be important, for example in investigating the dimensionless dissipation rate, whilst for others we will be more interested in the triadic interactions which are independent of the choice of $\lambda_1$. The forcing term $f(k)$ is chosen such that injected energy is distributed evenly across the forcing range of wave-numbers. This forcing allows for the energy dissipation, $\varepsilon$, to be set exactly at stationary state. As most of this work will be focused on quantities that are determined by the small scales of the flow, the exact form of forcing at the large scales is not an important choice. Moreover, we have made use of different forms of forcing and verified that the small scale behaviour is unaltered. 

A major advantage of this closure is that, since the energy spectrum is a smooth function of $k$, we can make use of a logarithmic discretisation of the wave-number space when solving numerically \citep{leith1971atmospheric}. This is the key feature that allows very high Reynolds numbers to be achieved. Throughout this work we discretise our numerical simulations with \begin{equation}
k_i = k_{\mathrm{min}} 2^{i/F}
\end{equation} and we choose $k_{\mathrm{min}} = 1$ and vary $F$ to maintain resolution. In particular, we choose $F$ such that the choice of $\lambda_1$ accurately determines the value of the Kolmogorov constant in the inertial range, see Appendices \ref{appA} and \ref{appB}. Due to the presence of oscillatory terms raised to the power of the spatial dimension in the EDQNM equation, a finer mesh, and thus higher value for $F$, is required.  Much consideration was given to this point by \citet{leith1972predictability} using the test field model. In this case they performed an averaging procedure on the oscillatory terms, using a geometric approach as opposed to the elegant analytic approach used by \citet{bowman1996wavenumber}. In order to simplify our approach we choose simply to use a finer mesh with increasing dimension rather than invoke an averaging procedure. If the simulation is under-resolved the value of the Kolmogorov constant will not be correct, as such, measurement of this constant provides a useful check for the accuracy of a simulation, we thus ensure that the value of the Kolmogorov constant measured in our simulations is in agreement with the results of Appendix \ref{appB}. 

It is important to note that it has been shown \citep{lesieur1978amortissement} that, when using such a logarithmic discretisation, interactions between certain non-local triads will not be accounted for in the EDQNM closure. It is possible to handle these missing triads analytically, however, for this study the extension of such methods to $d$-dimensions has not been carried out. Consequently, we have verified our main results are not influenced by the choice of discretisation through cross-checks with linearly discretised simulations where feasible. In our numerical work, we employ a parallel EDQNM code using second order predictor corrector method for 
time-stepping. Details of this code can be found in \citep{clark2019edqnm}.  

In order to determine an appropriate value for $\lambda_1$ in any dimension, we turn to results from a parameterless closure approximation. The Lagrangian renormalised approximation (LRA) \citep{kaneda1981renormalized} was used by \citet{gotoh2007statistical} to obtain values for the Kolmogorov constant in arbitrary dimension. Using these values we are then able to set the free parameter in our EDQNM simulations, such that we have an appropriate value for the Kolmogorov constant, see Appendix \ref{appB} for further details. In our numerical work above four dimensions $\lambda_1$ has been set such that the Kolmogorov constant in each dimension is equal to the value predicted by the LRA. Unfortunately, this ad-hoc method is all that is available as there are no DNS results beyond four dimensions from which a Kolmogorov constant can be approximated. 

In all simulations we start from a zero initial energy spectrum. Simulations are then evolved from this spectrum until a statistically stationary state is reached. The exact form of the initial condition is not important for this work: the statistically steady state is independent of such choices. In fact, as we begin from a zero state the steady state spectra is guaranteed to be generated entirely by the triadic interactions of equation \ref{edqnmd}. In Appendix \ref{appB} we give further details on the simulations performed included the largest Reynolds numbers reached.

\section{Results}\label{results}

\subsection{Energy and Transfer Spectra}

\begin{figure}
   \includegraphics{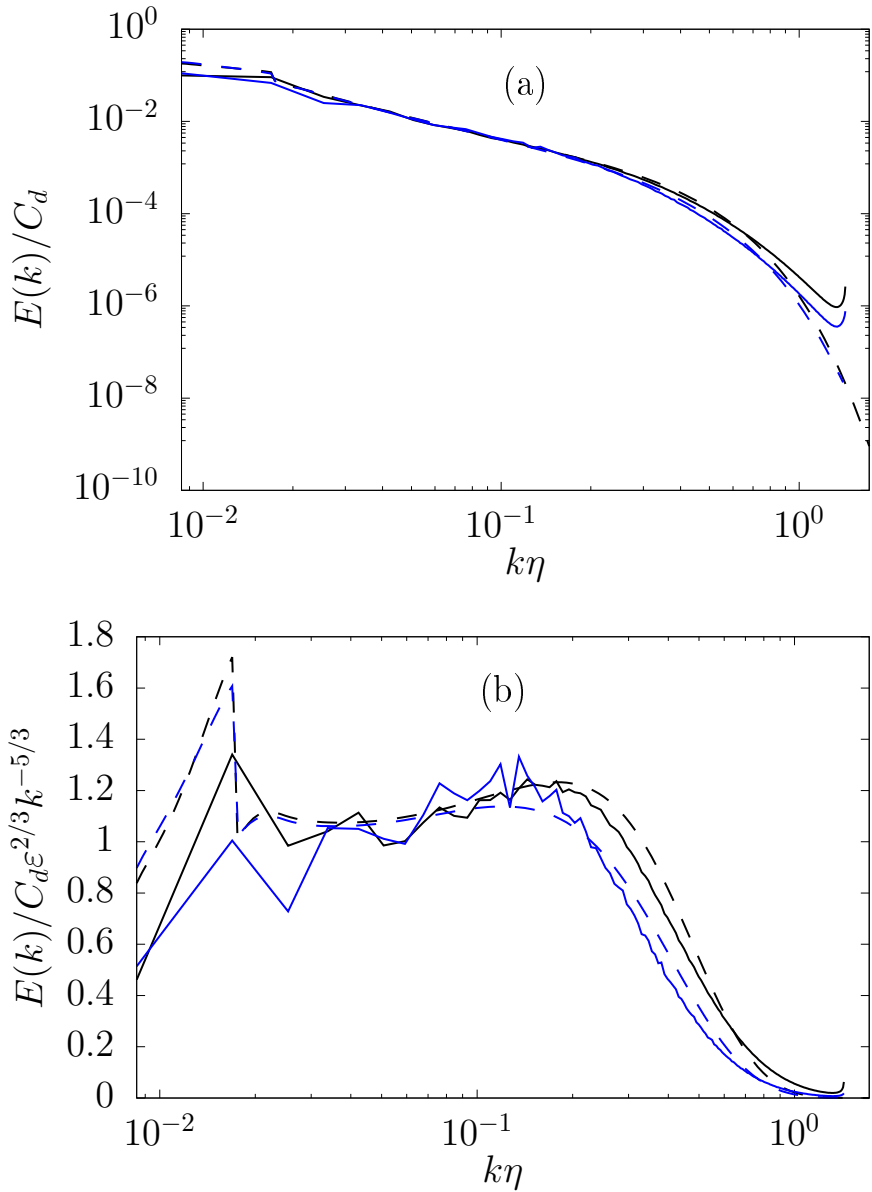}
    \caption{(a) Comparison between DNS (solid lines) and EDQNM (dashed lines) of energy spectra scaled by the Kolmogorov microscale, $\eta$, for three (blue) and four (black) dimensions. (b) Compensated energy spectra for the same data.}
     \label{dnsspec}
\end{figure}

In the DNS study of \citet{berera2019d}, the observed scaling of the energy spectrum was consistent with that predicted by K41 in both the three and four dimensional cases insofar as within a given dimension the energy spectra were found to collapse upon scaling by Kolmogorov variables. However, the spatial dimension was found to have an influence on the shape of the energy spectrum in the near dissipative 
region.  It was observed that, when comparing four to 
three dimensions, dissipative effects did not become dominant until smaller 
scales, evidenced by the presence of a seemingly extended inertial range. 
Before we investigate this behaviour in higher dimensions using the EDQNM 
closure, we need to understand to what extent the model is capable of reproducing the effects seen in three and four-dimensional DNS. To this end, in figure \ref{dnsspec} we show energy spectra from both DNS and EDQNM in three and four dimensions scaled by  the Kolmogorov constant. In both dimensions we have the same $\nu$ and $\varepsilon$ across DNS and EDQNM. A good collapse of the data can be seen in the inertial range and, approaching the dissipative region, the EDQNM closure captures the extended inertial range reasonably well. However, once the dissipative region is reached, the EDQNM results begin to diverge from those of the DNS in both 
dimensions, and it does not appear that this divergence is worse in one dimension over the other. The discrepancies at the large scales, small $k\eta$, are due to the forcing differences between DNS and EDQNM simulations.

\begin{figure}
    \includegraphics{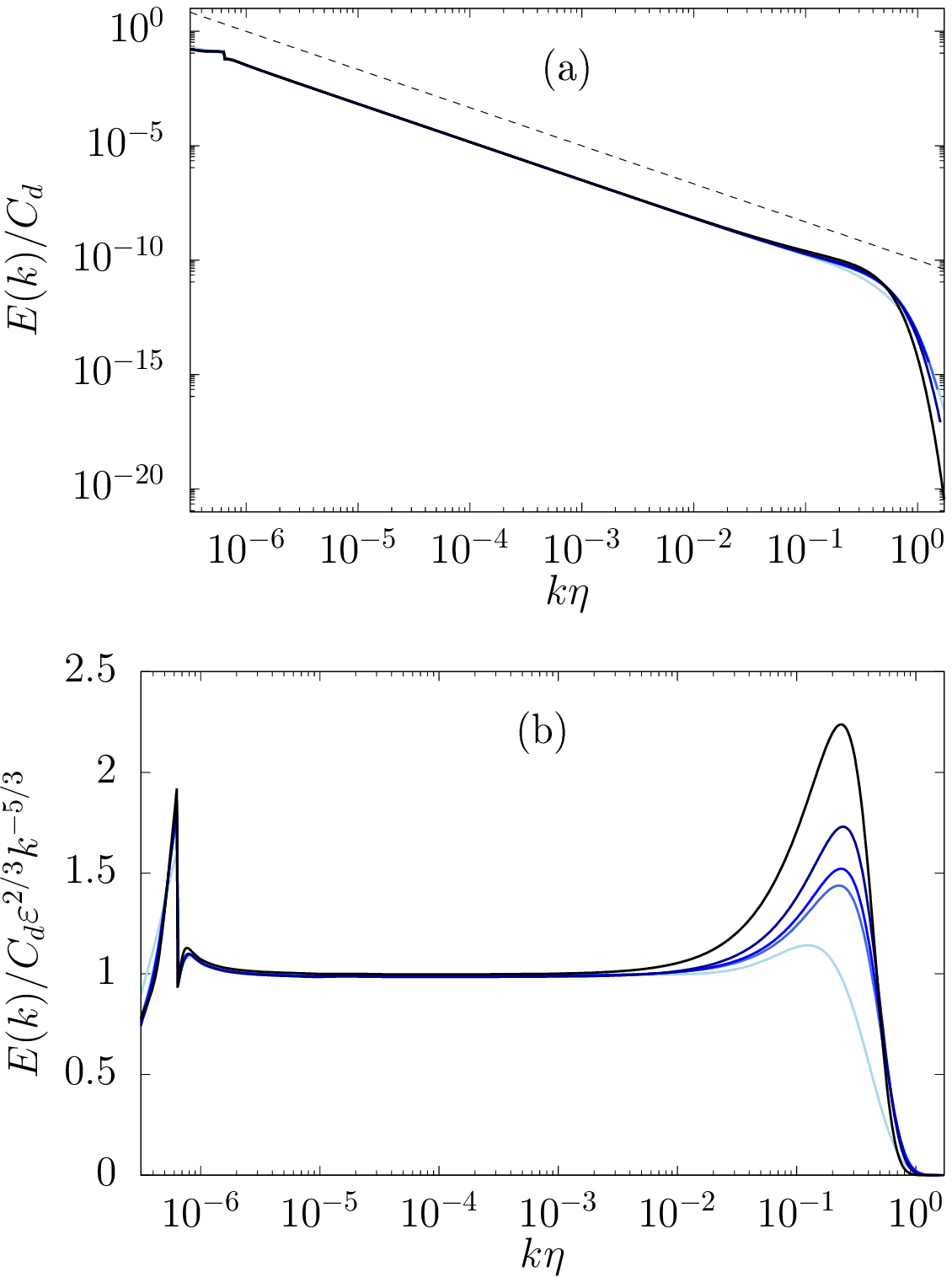}
    \caption{(a) EDQNM energy spectra scaled by the Kolmogorov microscale, $\eta$, for three, six, seven, ten and twenty dimensions, the darker the shade of the line the higher the dimension with twenty being black. Dashed line shows $k^{-5/3}$ scaling. All dimensions have the same viscosity and energy dissipation. (b) Compensated spectra for the same data, colours the same.}
    \label{spec}
\end{figure}

Having verified that the EDQNM approximation can satisfactorily reproduce properties of the energy spectra seen in three and four dimensional DNS, we now turn to even higher spatial dimensions. In figure \ref{spec}a we plot the energy spectra from EDQNM simulations for three, six, seven, ten and twenty dimensions, scaled by the appropriate Kolmogorov constant for each dimension, once more we keep $\nu$ and $\varepsilon$ constant across dimensions. Here, it can be seen that increasing spatial dimension is accompanied by a growing accumulation of energy on the edge of the inertial range. Such behaviour may be indicative of an enhanced forward transfer of energy as the spatial dimension increases. This view is consistent with theoretical arguments which conjecture that as the spatial dimension tends to infinity the nature of the triadic interactions leads to all energy being transferred in the forward direction to the small scales \citep{fournier1978infinite}. Further arguments have suggested that the appearance of such an energy bottleneck is the result of triad interactions being damped by viscosity at the smallest scales \citep{herring1982comparative}. Hence, if the forward energy transfer is enhanced and, as appears to be the case, the aforementioned viscous damping of certain triads is not influenced by the dimension, then this pile up should be expected to increase with dimension. In figure \ref{spec}b, we show the compensated spectra which give an even clearer demonstration that the bottleneck effect becomes more pronounced with increasing dimension. In DNS results \citep{berera2019d}, this view is further supported by an increased skewness in four dimensions.  We will return to skewness in section \ref{skewnss} where it is evaluated for EDQNM results. It is also clear that in all cases we observe a inertial range with a $k^{-5/3}$ power law scaling which persists over a number of decades in wave-number space. As we go to higher dimensions we find this scaling region appears to become progressively shortened by the increased bottleneck effect. Although not shown, in all dimensions we find a collapse within a given dimension of energy spectra across a range of $\Rey$ values when rescaled by $\nu$ and $\varepsilon$. That is, the energy spectrum is found to take on a universal shape in each dimension.

\begin{figure}
    \includegraphics{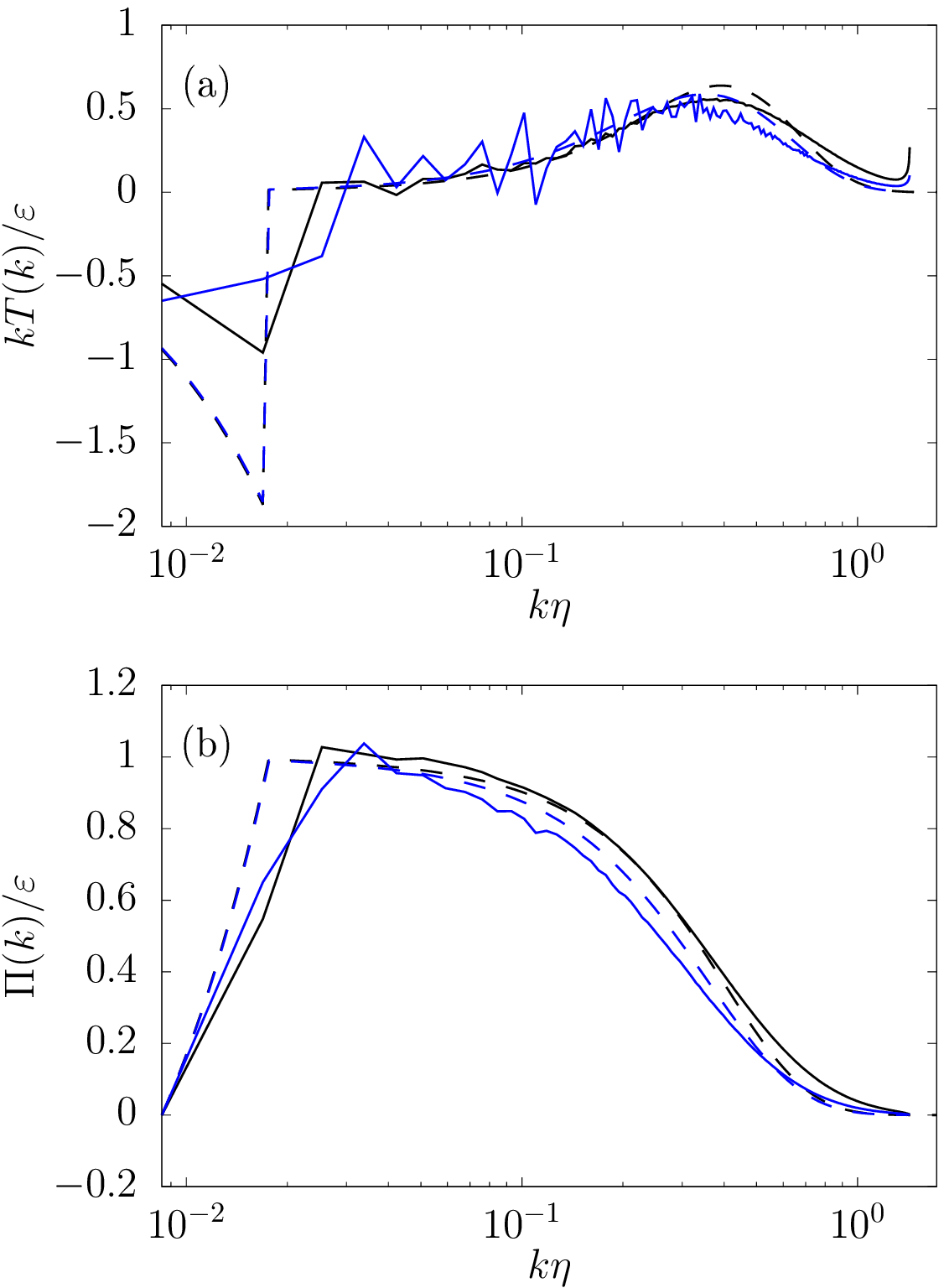}
    \caption{(a) Comparison between DNS (solid lines) and EDQNM (dashed lines) of non-linear transfer in three (blue) and four (black) dimensions. (b) Normalised energy flux for DNS (solid lines) and EDQNM (dashed lines) in three (blue) and four (black) dimensions.}
    \label{dnsnonlin}
\end{figure}

The energy transfer spectrum, $T(k)$, provides a natural measure of the exchange of energy between different scales in turbulent flows. However, in closing the infinite moment hierarchy using the quasi-normal hypothesis, the transfer spectrum is directly effected by the closure. Moreover, $T(k)$ is also directly influenced by the eddy-damping assumption. As such, if the quasi-normal and/or the eddy-damping assumptions are not sound we should expect the transfer spectra produced in the EDQNM simulations to show differences when compared to DNS transfer spectra. Indeed, in figure \ref{dnsnonlin}a we see far larger discrepancies between EDQNM and DNS for the non-linear energy transfer than we did in the corresponding energy spectra. However, the qualitative behaviour going from three to four dimensions is the same in both DNS and EDQNM results. The peak non-linear transfer is greater and found at smaller scales in four dimensions compared to three dimensions. However, in contrast with results for the energy spectra, the agreement between the non-linear transfer in DNS and EDQNM appears to be better in four dimensions than in three dimensions, insofar as the peak transfer occurs at similar scales in both DNS and EDQNM. Without higher dimensional DNS results we cannot say whether this better agreement between DNS and EDQNM is purely coincidental or in fact evidence that four dimensional turbulence is in effect more mean-field-like. In figure \ref{dnsnonlin}b, the energy flux is displayed. For both DNS and EDQNM the energy flux remains roughly constant until smaller scales in four dimensions relative to three dimensions. Once more, due to the forcing differences at larger scales there is a greater disagreement between DNS and EDQNM results for both the non-linear transfer and the energy flux. In light of these comparisons, we should be more cautious in our interpretation of results derived from the transfer spectrum in EDQNM.

\begin{figure}
    \includegraphics{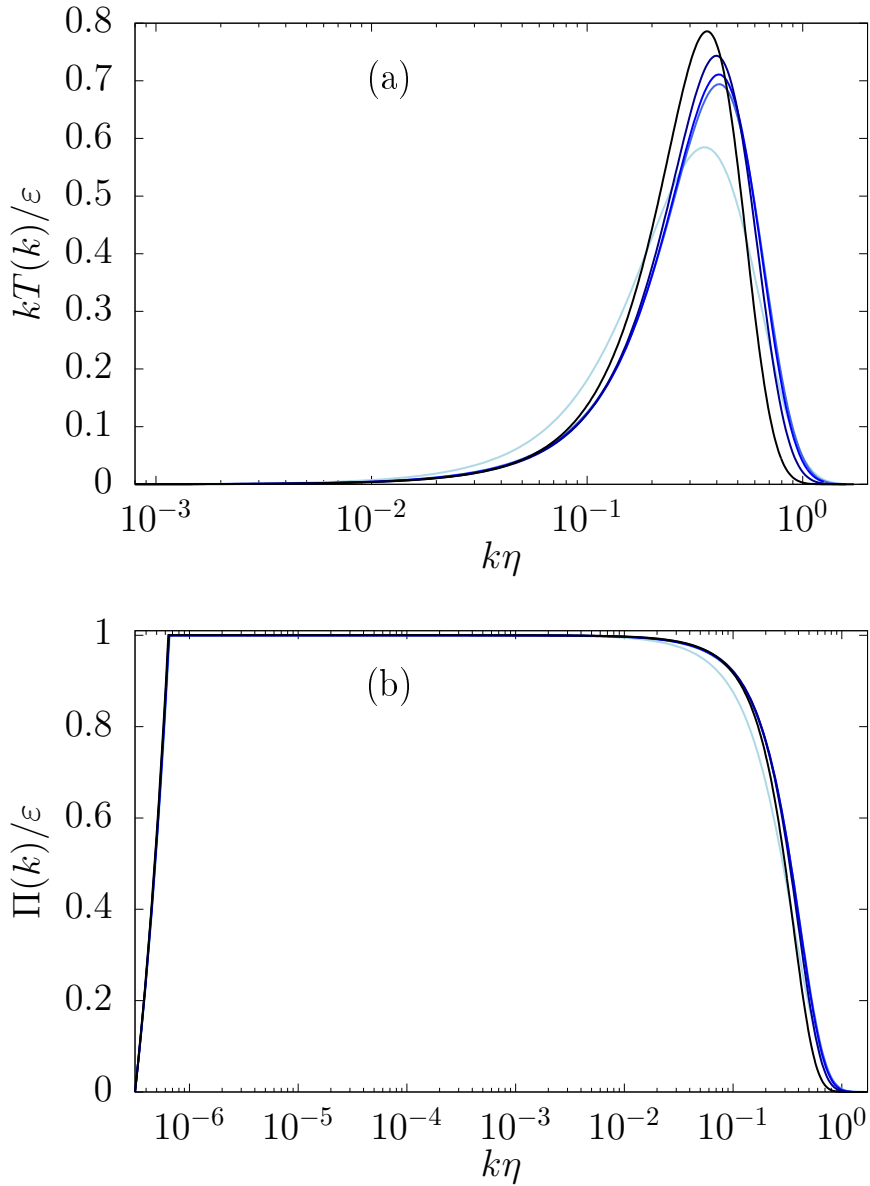}
    \caption{(a) EDQNM Non-linear energy transfer for three, six, seven, ten and twenty dimensions, the darker the shade of the line the higher the dimension with twenty being black. (b) Normalised energy flux for same data coloured in the same manner.}
    \label{nonlin}
\end{figure}

Turning once more to purely EDQNM results, we look at the dependence on the spatial dimension of the non-linear energy transfer. In figure \ref{nonlin}a we plot the non-linear energy transfer for a number of dimensions. Here, the trend of the peak non-linear energy transfer moving to smaller scales as the dimension increases is observed to continue to around six dimensions, at which point it begins to move to larger scales again. This potential crossover at six dimensions is interesting given the work of \citet{liao1990some, liao1991kolmogorov}, where a possible upper critical dimension for turbulence at six dimensions is conjectured. Of course, since our results are obtained via a closure approximation they should not be over-interpreted. In figure \ref{nonlin}b we show the spectral energy flux for a range of spatial dimensions. In line with what was found for the compensated energy spectra, we observe a scaling range of several decades in all dimensions. In this figure it is clear that above three dimensions there is an increased energy transfer to smaller scales as evidenced by the flux dropping slower as we enter the dissipative region.

\subsection{Skewness}\label{skewnss}

\begin{figure}
    \includegraphics{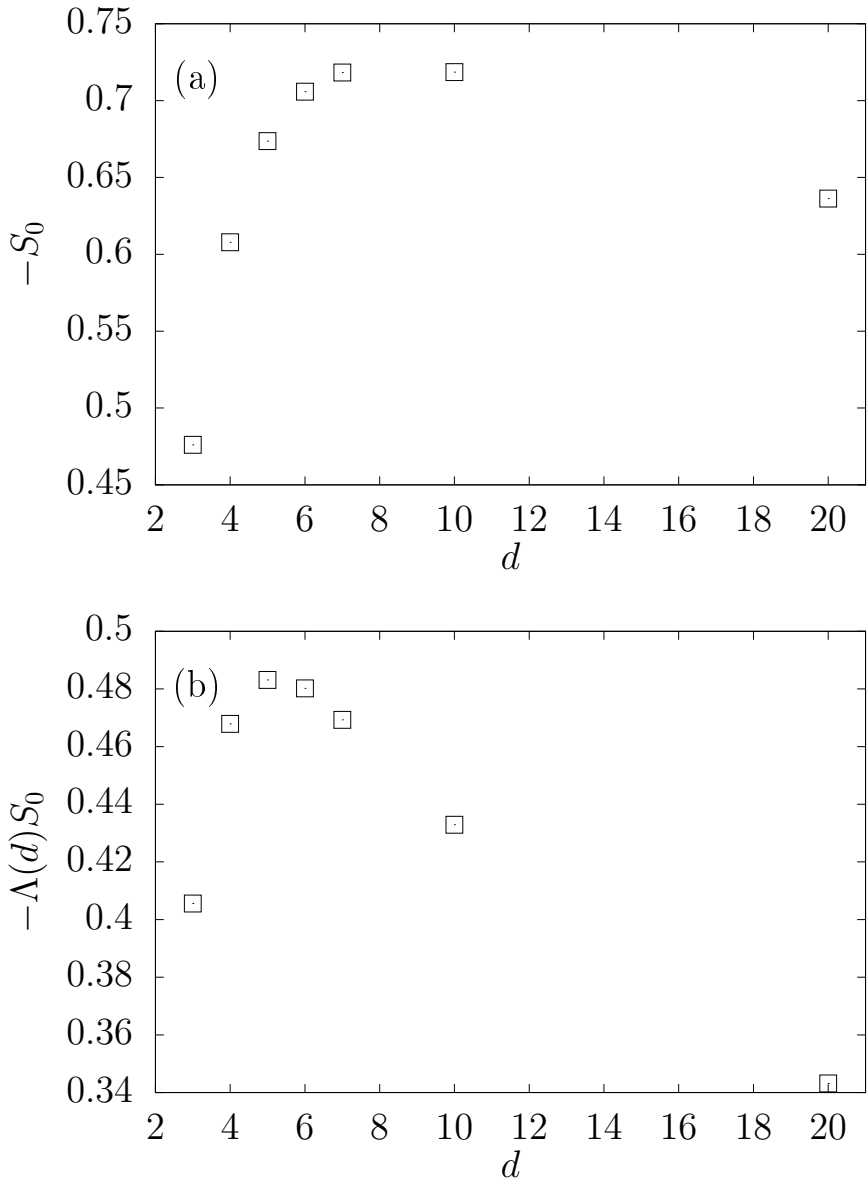}
    \caption{(a) EDQNM Velocity derivative skewness vs dimension. (b) Enstrophy production term $-S_{0} \Lambda(d)$ vs dimension.}
    \label{skew0}
\end{figure}

In section \ref{sectionskew} we derived equation (\ref{enstroskew}), relating the production of the generalised enstrophy in $d$-dimensions to the skewness, $S_0$. Higher values of $S_0$ are then associated with greater vortex-stretching, which would then provide the mechanism for the increased forward energy transfer in higher dimensions. Using equation (\ref{dskewness}) we can measure the effect of spatial dimension on $S_0$ in our EDQNM simulations. These results are presented in figure \ref{skew0}a. It is observed that in the EDQNM equations the skewness reaches a maximum value of around -0.72 at seven dimensions, before remaining roughly constant until ten dimensions, beyond which $S_{0}$ decreases. If the trend seen in figure \ref{skew0}a continues then the skewness may vanish for infinite spatial dimension. When compared to the results for skewness in the DNS of \citet{berera2019d} both three and four-dimensional EDQNM results exhibit a lower value of $S_0$. This is not a surprising result and has been observed in EDQNM simulations in three dimensions \citep{bos2012reynolds} and is likely a result of the assumptions made in the EDQNM model \textit{e.g.} the quasi-normal or eddy-damping assumptions. The EDQNM approximation is also known to exhibit a constant asymptotic value for $S_0$ at sufficiently high Reynolds number and the results presented in figures \ref{skew0}a and \ref{skew0}b are of this asymptotic value in all cases. This existence of this asymptotic value is predicated on the exponent in the inertial range being $-5/3$ hence given our energy spectra results in all dimensions we can be confident in these asymptotic values. 

In equation (\ref{enstroskew}), due to the dimensional pre-factor of the skewness term, at larger spatial dimensions higher skewness values are required to generate the same level of enstrophy production. In figure \ref{skew0}b we show the pre-factor, $-S_{0}\Lambda(d)$, of the enstrophy production term in equation (\ref{enstroskew}) for a range of dimensions. These results suggest that until five dimensions there is an increase in enstrophy production, then starting at six dimensions there is a reduction in enstrophy production and thus in vortex stretching. If we consider that the action of vortex stretching produces smaller scales in the flow, then this is consistent with the accumulation of energy at the end of the inertial range. Once more, if this trend continues, then for infinite dimension there will be no enstrophy production and hence no vortex stretching. The vanishing of vortex stretching, and thus of velocity derivative skewness, at infinite dimension is quite an extreme scenario, and it may be that a non-zero but finite asymptotic value is reached instead. This is consistent with what was seen in \citep{gotoh2007statistical} where a finite but non-zero asymptotic skewness value at infinite dimension was predicted using the LRA. This finite value is reached after a consistent increase with dimension. This is at odds with what is seen in our simulations, where a maximum skewness is reached at a finite dimension. However, if the skewness does vanish at infinite dimension a more drastic statement may then be that in infinite dimensions there is no energy flux in either direction and thus no turbulence. We stress here that since these are closure results it is not possible to make any definitive claims.

The significance of this maximum skewness dimension is unknown, and may be a result of the closure assumptions. As such, without higher dimensional DNS results, we cannot make a definitive statement. Furthermore, to ensure that the skewness reaching a maximum at finite dimension is not a result of the values of free parameter, $\lambda_1$, used in this work we have carried out an additional set of simulations detailed in Appendix \ref{appC}. As a result of these simulations we are confident the appearance of a maximum skewness dimension is a real effect in the EDQNM equations, with the caveat that the Kolmogorov constant does not increase with dimension. Given the observed Kolmogorov constants in the four dimensional DNS of both \citet{gotoh2007statistical} and \citet{berera2019d} and the predictions of the LRA there is evidence that this caveat holds.

An important further point to consider for the infinite dimensional problem is that of the energy spectrum normalisation in this limit. In \citep{fournier1978infinite} it was found that to each order in perturbation theory the energy spectrum had a finite limit if a rescaled time variable is used. However, this time rescaling can be shown to be equivalent to a rescaling of the energy spectrum. The latter rescaling is equivalent to a finite energy per velocity component as opposed to a finite total energy. If the total energy is to remain finite then as we tend to infinite dimension each individual velocity component will tend to zero, suggesting zero skewness. However if the components remain non-zero then the skewness will be finite but non-zero.

Once more, these results are interesting considering the work of \citet{liao1990some,liao1991kolmogorov} suggesting the possibility of a critical dimension of 6 for turbulence. Studying this infinite dimensional limit within numerical EDQNM studies becomes difficult with increasing dimensions. For example, at $d=20$ with a numerical resolution of $F=50$ we find an error of about 2\% in the expected value for the Kolmogorov constant using the free parameter in Appendix \ref{appB}. Compared with $d=3$ where a resolution of $F=16$ gives an error much less than a percent it is clear that the resolution costs of higher dimensions quickly becomes an issue.

\subsection{Third Order Structure Functions}

The longitudinal structure functions have frequently been measured in experimental \citep{anselmet1984high} and numerical studies of turbulence \citep{gotoh2002velocity,ishihara2009study}. In such studies, it is found that, particularly at higher orders, these structure functions show deviations from the scaling predicted by the K41 theory. Such deviations are typically attributed to intermittency corrections, although there are also arguments suggesting these are simply finite Reynolds number corrections due to K41 being an asymptotic theory \citep{kolmogorov1962refinement, benzi1984multifractal, qian1997inertial, qian1999slow, antonia2006approach, tchoufag2012spectral, mccomb2014homogeneous, tang2017finite, antonia2019finite}. 

The effect of the spatial dimension on these corrections is an interesting question, and comparisons to critical phenomena, in particular anomalous exponents, have been made \citep{nelkin1974turbulence,rose1978fully}. Of course, the EDQNM model does not account for intermittency and therefore measurements of structure functions in such simulations cannot answer questions regarding these deviations. Furthermore, we are not aware of a method by which to calculate beyond third order structure functions from spectral quantities. However, if such a method were available it is likely it could, potentially with some difficulty, be applied to the EDQNM model. Indeed, it was shown by \citet{kraichnan1959structure} that in the Direct Interaction Approximation (DIA) it is possible to study quantities beyond third order. This was then demonstrated by \citet{chen1989non} for the DIA and by \citet{bos2013strength} for the EDQNM model. As such, even if intermittency was present in the EDQNM model we would be unlikely to find substantial corrections at these low orders, even in three dimensions.

\begin{figure}
    \includegraphics{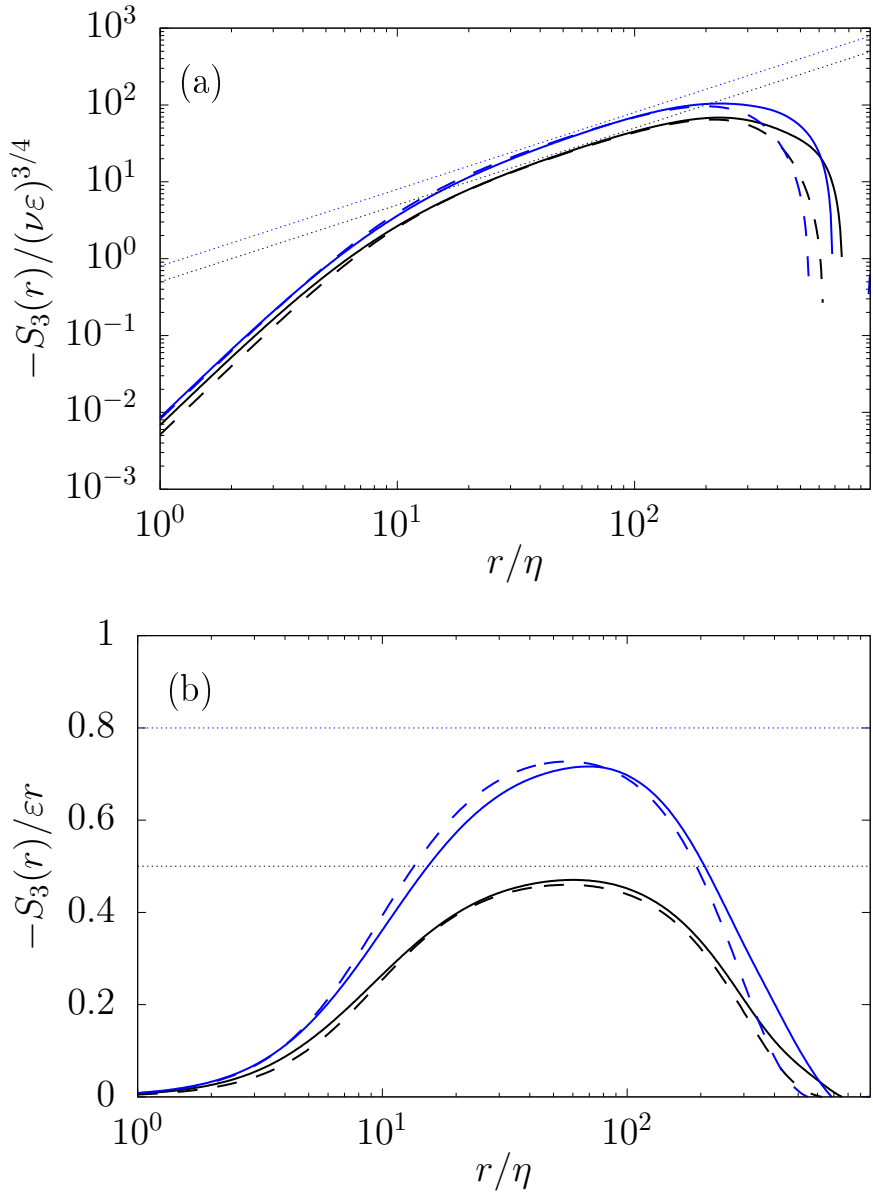}
    \caption{Comparison of third order structure functions in DNS (solid lines) and EDQNM (short dashed lines) for three (blue) and four (black) dimensions: (a) scaled by appropriate Kolmogorov quantities (b) scaled by energy dissipation and $r$. In (a) the dotted lines correspond to the expected $12r/d(d+2)$ inertial range scaling behaviours whilst in (b) they are for the values $12/d(d+2)$.}
    \label{struct3dns}
\end{figure}

\begin{figure}
    \includegraphics{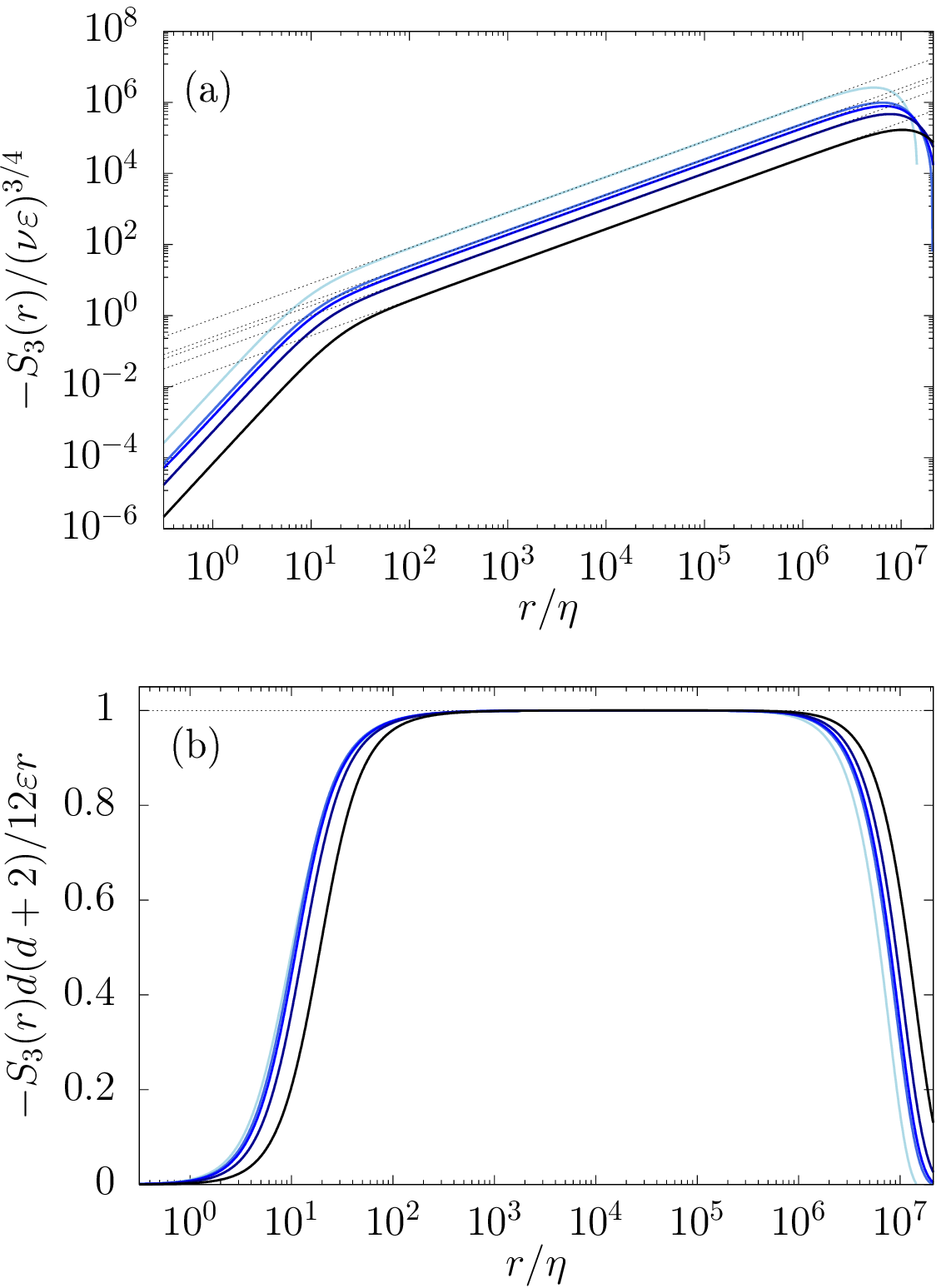}
    \caption{EDQNM Third order structure functions scaled by appropriate Kolmogorov quantities for three, six, seven, ten and twenty dimensions, the darker the shade of the line the higher the dimension with twenty being black. Dashed lines represent appropriate powerlaw scaling for each dimension, \textit{i.e.} $12r/d(d+2)$ for the inertial range. (b) Third order structure function scaled by energy dissipation and $r$ for the same data with the same colouring.}
    \label{struct3}
\end{figure}

Putting questions regarding intermittency and anomalous exponents aside, 
we wish to test equation (\ref{3struct}) using our DNS and EDQNM results. 
In figures \ref{struct3dns}a and \ref{struct3dns}b, the third 
order structure functions computed using equation (\ref{3struct}) in EDQNM 
are compared
to those in DNS.  Good agreement is seen in the inertial range with 
both DNS and EDQNM exhibiting the expected scaling. Looking to the 
dissipative region, we find that all our data follows $r^3$  scaling, 
however, this scaling begins at a different point in DNS compared to EDQNM. 
Given our energy and transfer spectra results, these deviations in small 
scale behaviour are expected. The most interesting feature of these figures can be found in figure \ref{struct3dns}b where the agreement between DNS and EDQNM appears to be better in four dimensions when compared with three. This is consistent with what was seen in the non-linear transfer, perhaps not surprisingly given that the transfer spectrum is used in determining $S^{(d)}_{3}(r)$. Without higher dimensional DNS data it is impossible to know if this better agreement is due to the EDQNM approximation becoming more accurate with increasing dimension or simply a coincidence. 

As already stated, the EDQNM model does not exhibit intermittency. However, it does capture well the finite Reynolds number effect. Indeed, in \cite{bos2012reynolds} the EDQNM model was compared with the multi-fractal model for three dimensional turbulence giving comparable results for low $\Rey$ suggesting it is difficult to distinguish between intermittency effects and finite Reynolds effects in this region. We have not studied the multi-fractal model in four dimensions to make this comparison, however, given the better agreement between EDQNM and DNS in four dimensions for $S^{(d)}_{3}(r)$ it may be the case that the finite Reynolds number effect becomes dominant over intermittency in higher dimensions.

 Looking now at figure \ref{struct3}a, we see 
the third order lognitudinal structure function for our EDQNM data in 
four, five and six dimensions scaled by appropriate Kolmogorov quantities. 
It can be seen that in all dimensions we have $r^3$ scaling in the 
small $r$ limit, as is seen in three dimensional turbulence and is predicted from the small $r$ expansion of $S^{(d)}_{3}(r)$. 
Turning to the $d$-dimensional von K\'{a}rm\'{a}n-Howarth equation, 
it can be shown that in the inertial range for the third order structure 
function, we should find \begin{equation}
S_{3}^{(d)}(r) \simeq -\frac{12}{d(d+2)}\varepsilon r \;,
\end{equation} which reduces to the standard four-fifths law of three dimensional turbulence. Indeed, in figure \ref{struct3}a we can see that each dimension follows its own $-12/d(d+2)$ law in the inertial range. Once more in all dimensions we observe a long scaling region. For a clearer comparison in figure \ref{struct3}b we normalise each dimension by the expect inertial range value such that all cases show an scaling range at 1. In doing so we find differences across dimensions, in particular by $d=20$ the scaling region begins at higher $r/\eta$ than in lower dimensions.

\subsection{Dissipative Anomaly}

In both experimental \citep{sreenivasan1984scaling, burattini2005normalized} and numerical studies \citep{wang1996examination, kaneda2003energy, ishihara2016energy}  of three dimensional turbulence, including in EDQNM \citep{bos2007spectral}, there is a large body of evidence which indicates the existence of a non-zero energy dissipation rate, even in the limit of zero viscosity. This is known as the dissipative anomaly. In \citep{berera2019d} an increased value for this asymptotic dissipation rate  was observed in four dimensions when compared with three. This result once again suggests an enhancement of the forward energy cascade in four dimensions compared with three. We should then expect that beyond four dimensions this asymptotic dissipation rate should increase further given our spectral and skewness results. 

The dimensionless dissipation rate is defined as \begin{equation}
C_{\varepsilon} = \frac{\varepsilon L}{u^3},
\end{equation} and its Reynolds number dependence can be shown to be approximately described by the relationship \citep{doering2002energy, mccomb2015energy} \begin{equation}\label{cepseqn}
C_{\varepsilon}(\Rey) = C_{\varepsilon, \infty} + \frac{C}{\Rey} \;,
\end{equation} where $\Rey$ $=uL/\nu$ is the integral scale Reynolds number in which $L$ is defined by equation (\ref{L}). 

\begin{figure}
    \includegraphics{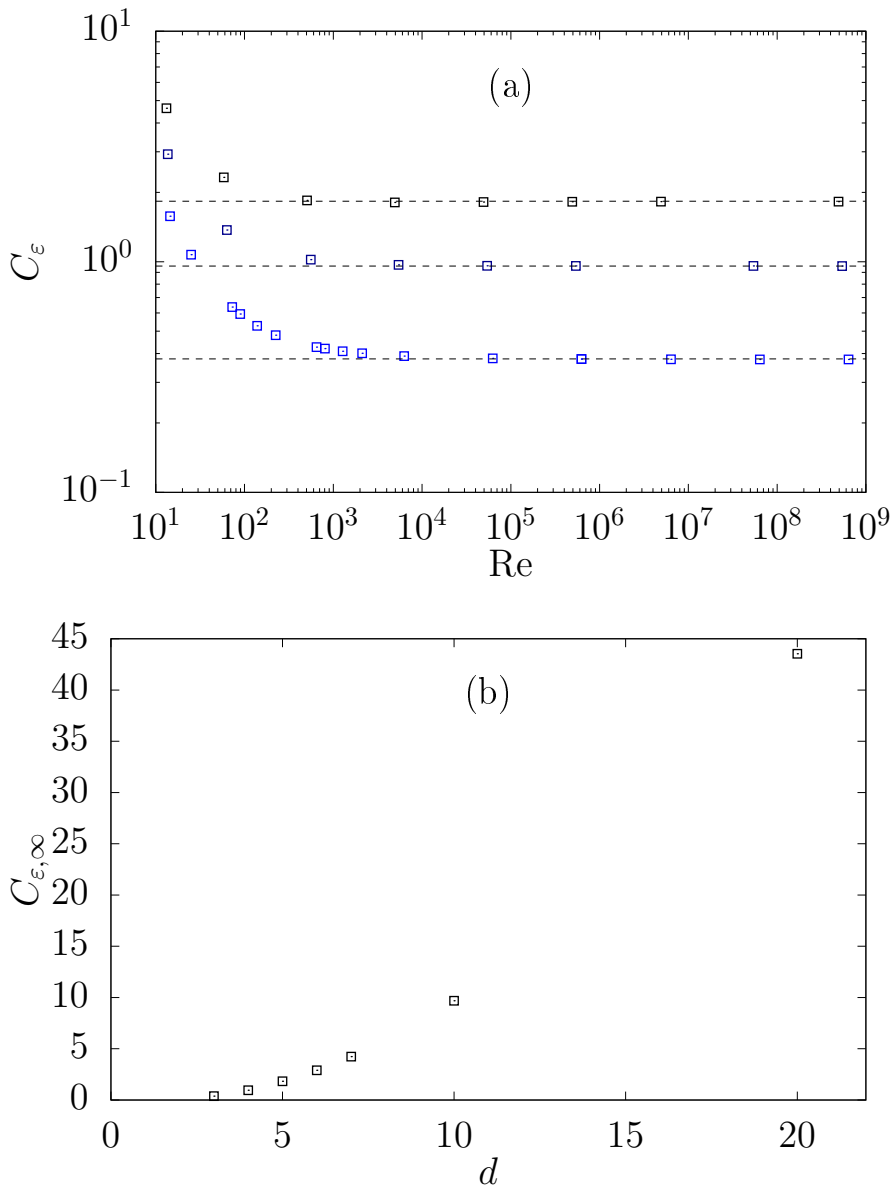}
    \caption{a) $C_{\varepsilon}$ vs $\Rey$ for three, four and five dimensions, darker colour indicates higher dimension with five dimensions in black. Dashed lines indicate the value for $C_{\varepsilon, \infty}$. b) $C_{\varepsilon, \infty}$ against spatial dimension, $d.$}
    \label{ceps}
\end{figure}

In figure \ref{ceps}a we show the dimensionless dissipation rate against $\Rey$  for a wide range of $\Rey$ values for three, four and five dimensions. In all cases we observe $C_{\varepsilon}$ tending to a constant asymptotic value. This asymptotic value is seen to grow with dimension as would be expected from increased forward energy transfer. We find that in three dimensions this asymptotic value in EDQNM is $C^{3d}_{\varepsilon,\infty} = 0.38$ lower than what is seen in DNS where $C^{3d}_{\varepsilon,\infty} \approx 0.5$ in the forced case. A higher value was found in the EDQNM of \citet{bos2007spectral} however they take a different choice of $\lambda_1$ which will directly influence the value found. In four dimensions we find $C^{4d}_{\varepsilon,\infty} = 0.96$ lower than the value of 1.26 found in DNS \citep{berera2019d}. Since $C_{\varepsilon}$ is defined in terms of large scale quantities, where the forcing is active, these discrepancies between DNS and EDQNM are not surprising given the differences at small $k$ seen in figure \ref{dnsnonlin}. In figure \ref{ceps}b we show the asymptotic dimensionless dissipation rate against the spatial dimension. Here we find the asymptotic value grows with dimension until at least 10 dimensions. Beyond here the resolution requirements render further calculation increasingly difficult. As $C_{\varepsilon, \infty}$ is related to the Kolmogorov constant such measurements would become exceptionally sensitive to resolution errors.

\section{Conclusion}\label{conclusion}
 
Motivated by the four dimensional DNS results presented 
in \citet{berera2019d}, 
as well as analogies to critical phenomena, 
in this paper we have performed a thorough investigation into the effect 
of the spatial dimension in the EDQNM model of turbulence. While this is 
only an approximation to true fluid turbulence, we find that it is able to 
satisfactorily reproduce many of the dimensional effects seen in DNS. 
To facilitate this study, a number of standard results from three dimensional 
turbulence have been extended to $d$-dimensional turbulence. 
Some of these quantities have been discussed in the literature with
theoretical ideas as to how they will behave in higher dimensions. This paper has presented for the first time both
numerical results using EDQNM in a range of dimensions above three, and
expressions for the second and third order structure function
in terms of spectral quantities for any dimension $d$. Furthermore, an equation relating the production of enstrophy to skewness in $d$-dimensional turbulence was derived from the von K\'{a}rm\'{a}n-Howarth equation.

The energy and transfer spectra were measured across a wide range of spatial dimensions. For the three and four dimensional cases comparisons were made with DNS results where it was observed that the EDQNM model accurately captures the dimensional differences seen in DNS. In these measurements along with those of the skewness
as function of spatial dimension, we have found a consistent
picture suggesting the forward energy cascade becomes enhanced as 
spatial dimension increases.
In terms of spectra, this can be seen as an increase in the bottleneck effect in the near dissipation region of the energy spectra, which grows with dimension. This suggests that the viscous damping of triad interactions is either not effected by dimension, or is enhanced at a lesser rate than the energy transfer. 
Further to this, in the transfer spectra, we see a larger peak in the 
non-linear transfer. We also observe that the position of this peak 
first moves to smaller scales before then reversing and moving back to 
larger scales, which we posit is a result of suppression of transfer to the 
dissipative modes as suggested by \citet{herring1982comparative}. We find 
that enstrophy production reaches a maximum in five dimensions, possibly going to zero at very high dimension. In light of the enstrophy-skewness equation we have derived, 
this corresponds to a reduction in small scale vortex stretching, which 
is again consistent with the enhanced forward transfer bottleneck effect. 
The possibility of a zero enstrophy production in the limit of infinite 
dimensions poses interesting questions for the fate of turbulence 
in this limit.

Additionally, we have measured the third order structure functions in 
higher dimensions using the spectral relations we have derived. We find 
here that each dimension has its own analogue to the four-fifths law of 
three dimensional turbulence. Interestingly, when looking at the third order structure function, the comparison between EDQNM and DNS appears better in four dimensions compared with three suggesting possible changes in the turbulent dynamics. Finally, we studied the effect of the spatial 
dimension on the asymptotic dissipation rate. Due to being defined in 
terms of large-scale quantities, this is more difficult to accurately 
measure in the EDQNM model. However, we do find an increase in this 
asymptotic dissipation rate with dimension, which is a continuation of 
the trend seen in the DNS performed in \citet{berera2019d}. This is also consistent with the existence 
of an enhanced forward transfer of energy in higher dimensions.

These results are interesting for a number of reasons.
Importantly, they confirm many of the results found in four spatial
dimensions from DNS \citet{berera2019d}.  The fact the EDQNM results
show consistency with DNS in three and four dimensions then suggests
the trends found by this method at even higher dimensions, which
at present are computationally too demanding for DNS, 
should have some reliability.  Thus, this paper has helped
to examine some of the theoretical ideas that have been in
the literature for decades on the behaviour of turbulence
in dimensions greater than three.
Furthermore, the appearance of an 
increased bottleneck effect in higher dimensions may help to shed light 
on the standard three dimensional bottleneck effect. Perhaps counter-intuitively, these results are also of interest for how \textit{little} turbulence changes from three to twenty dimensions. Indeed, compared with the dramatic changes observed going from three to two dimensions the differences between three and twenty are minimal, which is certainly intriguing.

In \citet{berera2019d} the scaling behaviour 
of fluctuations with Reynolds number was measured and found to decrease 
in four dimensions compared with three. In this work, temporal fluctuations in the total energy were measured and found to scale slower with $\Rey$ in four dimensions. It would be interesting to understand how such fluctuations would scale in even higher dimensions, especially in light of the results in this paper concerning the bottleneck effect and velocity derivative skewness. However, measurement of these fluctuations is unfortunately outside the scope of EDQNM calculations and 
would require future DNS study.

\section*{Acknowledgements}

This work has used resources from ARCHER (http://www.archer.ac.uk) via
the Director’s Time budget. D.C. is supported by the University
of Edinburgh, R.D.J.G.H is supported by the U.K.
Engineering and Physical Sciences Research Council
(EP/M506515/1). A.B. acknowledges funding from the U.K.
Science and Technology Facilities Council.

\section*{Declaration of Interests}
The authors report no conflict of interest

\appendix
\section{Setting the Kolmogorov Constant}\label{appA}

In order to make use of the EDQNM model, it is necessary to specify the value of the free parameter $\lambda_1$ seen in equation \ref{edqnmd}. The value of this constant can be shown to fix the value of the Kolmogorov constant, $C_d$, which is important for a number of numerical measurements. Here, we extend the method used by \citet{mccomb1990physics} to derive of the relationship between $\lambda_1$ and $C_d$ to the $d$-dimensional case. An alternative derivation of the relationship for the three dimensional case can also be found in \cite{andre1977influence}.

To begin, we consider the eddy-damping rate defined in equation \ref{eddamp} in the limit $\nu \rightarrow 0$, such that the energy dissipation rate remains constant and,  taking the energy spectrum to be a Kolmogorov spectrum to infinity and find \begin{equation}
\mu_{k} = \frac{\sqrt{3}}{2}\lambda_1 \sqrt{C_d} \varepsilon^{\frac{1}{2}}k^{\frac{2}{3}}\;.
\end{equation} On dimensional grounds, in the inertial range we can take $\mu_{k}$ to have the form \begin{equation}
\mu_{k} = \beta_d \varepsilon^{\frac{1}{2}}k^{\frac{2}{3}}\;,
\end{equation} and thus we have \begin{equation}\label{betaeq}
C_{d} = \frac{4\beta^{2}_{d}}{3\lambda^{2}_{1}}\;.
\end{equation} 
To make further progress we now need to relate $C_{d}$ and $\beta$ in the EDQNM model. We begin by considering the forced Lin equation at stationary state \begin{equation}
-T(k) = -2\nu k^2 E(k) + F(k)\;,
\end{equation} where $F(k)$ is the forcing spectrum. Now, in taking the limit of zero viscosity, the dissipation range will move to the smallest possible scales, hence we have \begin{equation}
2\nu k^2 E(k) \rightarrow \varepsilon\delta(k-\infty)\;,
\end{equation} and energy conservation then implies that \begin{equation}
F(k) \rightarrow \varepsilon\delta(k)\;.
\end{equation} Upon integrating both sides of the Lin equation up to a value $\kappa$, which by the symmetry properties of the integral is arbitrary so long as it is neither 0 nor $\infty$, we find \begin{equation}
-\int_{0}^{\kappa} dk \, T(k) = \varepsilon\;.
\end{equation} In the EDQNM model from equation \ref{edqnmd}, we have a closed expression for $T(k)$, using which we can perform this integration and relate $C_d$ and $\lambda_1$. Hence, we have \begin{equation}
\begin{split}
8K_d \int_{0}^{\kappa} \,  dk \iint\limits_{\Delta k} \mathrm{d}p \, \mathrm{d}q \, &\frac{k}{pq}b^{(d)}_{kpq}\theta_{kpq} \\ &\times \bigg[\sin^{d-3}(\beta) p^2 E(q)E(k)- \sin^{d-3}(\alpha) k^2 E(p)E(q)\bigg] = \varepsilon\;.
\end{split}
\end{equation} We can re-express the above integral as \begin{equation}
\begin{split}
-8K_d C^{2}_{d}\int_{0}^{\kappa} dk \, \int_{\kappa}^{\infty} dp \, \int_{|k-p|}^{k+p} dq \,  &\frac{k}{pq}b^{(d)}_{kpq} \\ &\times \frac{\sin^{d-3}(\beta) p^2 q^{-\frac{5}{3}}k^{-\frac{5}{3}} - \sin^{d-3}(\alpha) k^2 p^{-\frac{5}{3}}q^{-\frac{5}{3}} }{\beta_d\left(k^{\frac{2}{3}} + p^{\frac{2}{3}} + q^{\frac{2}{3}}\right)}= 1\;.
\end{split}
\end{equation} The two sine terms can also be expressed in terms of $k,p$ and $q$ as \begin{equation}
\sin (\alpha) = \sqrt{1 - \left(\frac{p^2 + q^2 - k^2}{2pq}\right)^2} \quad \mathrm{and} \quad \sin (\beta) = \sqrt{1 - \left(\frac{k^2 + q^2 - p^2}{2kq}\right)^2}\;.
\end{equation} The resulting integral must be evaluated numerically and, if we denote it by $I_d$ then we have \begin{equation}
\begin{split}
I_d = \int_{0}^{\kappa} dk \, \int_{\kappa}^{\infty} dp \, \int_{|k-p|}^{k+p} dq \,  &\frac{k}{pq}b^{(d)}_{kpq} \\ &\times \frac{\sin^{d-3}(\beta) p^2 q^{-\frac{5}{3}}k^{-\frac{5}{3}} - \sin^{d-3}(\alpha) k^2 p^{-\frac{5}{3}}q^{-\frac{5}{3}} }{\left(k^{\frac{2}{3}} + p^{\frac{2}{3}} + q^{\frac{2}{3}}\right)}\;,
\end{split}
\end{equation} and \begin{equation}
\frac{C^{2}_{d} I_d}{\beta_d} = \frac{1}{8K_d}\;.
\end{equation}  Then using equation \ref{betaeq} we can eliminate $\beta_d$, after which we find \begin{equation}
C_d = \left(\frac{\sqrt{3}}{16 K_d I_d} \right)^{\frac{2}{3}}\lambda^{\frac{2}{3}}_{1}\;.
\end{equation} This is our desired result and after numerical evaluation of $I_d$ this can be used to fix $C_d$ in simulation.

\begin{table}\label{tabA}
  \begin{center}
     \begin{tabular}{lllll}
\hline
$d$ & $C_d$ & $\lambda_1$ & Re$_L$ & Re$_\lambda$ \\ \hline
3   & 1.72  & 0.49  &  $6.4 \times 10^8$ & $1.6\times10^5$       \\
4   & 1.33  & 0.366 &  $5.4 \times 10^8$ & $1.2\times10^5$       \\
5   & 1.16  & 0.28  &  $4.9 \times 10^8$ & $1 \times10^5$      \\
6   & 1.08  & 0.23  &  $4.7 \times 10^8$ & $8.8\times10^4$       \\
7   & 1.03  & 0.195 &  $4.5 \times 10^8$ & $8.2\times10^4$             \\
10  & 0.952 & 0.134  &  $4.2 \times 10^8$ & $7.2\times10^4$     \\
20  & 0.877 & 0.066 &  $4 \times 10^8$ & $6.3\times10^4$       \\ \hline
  \end{tabular}
  \caption{Simulation parameters for all dimensions. In all cases we have $\nu = 1 \times 10^{-9}$ and $\varepsilon = 0.1$. Here, Re$_L$ is the integral scale Reynolds number and Re$_\lambda$ is the Taylor Reynolds number.}
  \end{center}
\end{table}

To make this more concrete we present the results of this procedure for the cases $d=3$ and $4$ where we take $\kappa=1$. For $d=3$ the dimensional factor $K_d$, which results from performing spherical integration in $d$-dimensions, is $K_3 = 1/8$, hence, we look to solve \begin{equation}
C_3 = \left(\frac{\sqrt{3}}{2 I_d} \right)^{\frac{2}{3}}\lambda^{\frac{2}{3}}_{1}\;.
\end{equation} It is found that $I_d \approx 0.19038$ and therefore for $d=3$ we have \begin{equation}
C_3 \approx 2.75 \lambda^{\frac{2}{3}}_{1}\;.
\end{equation} This  pre-factor differs very slightly compared with \cite{andre1977influence}, however this is likely a result of different methods of numerical integration. Following a similar procedure for $d=4$ it is found that \begin{equation}
C_4 \approx 2.6 \lambda^{\frac{2}{3}}_{1}\;.
\end{equation}

\begin{figure}
    \includegraphics{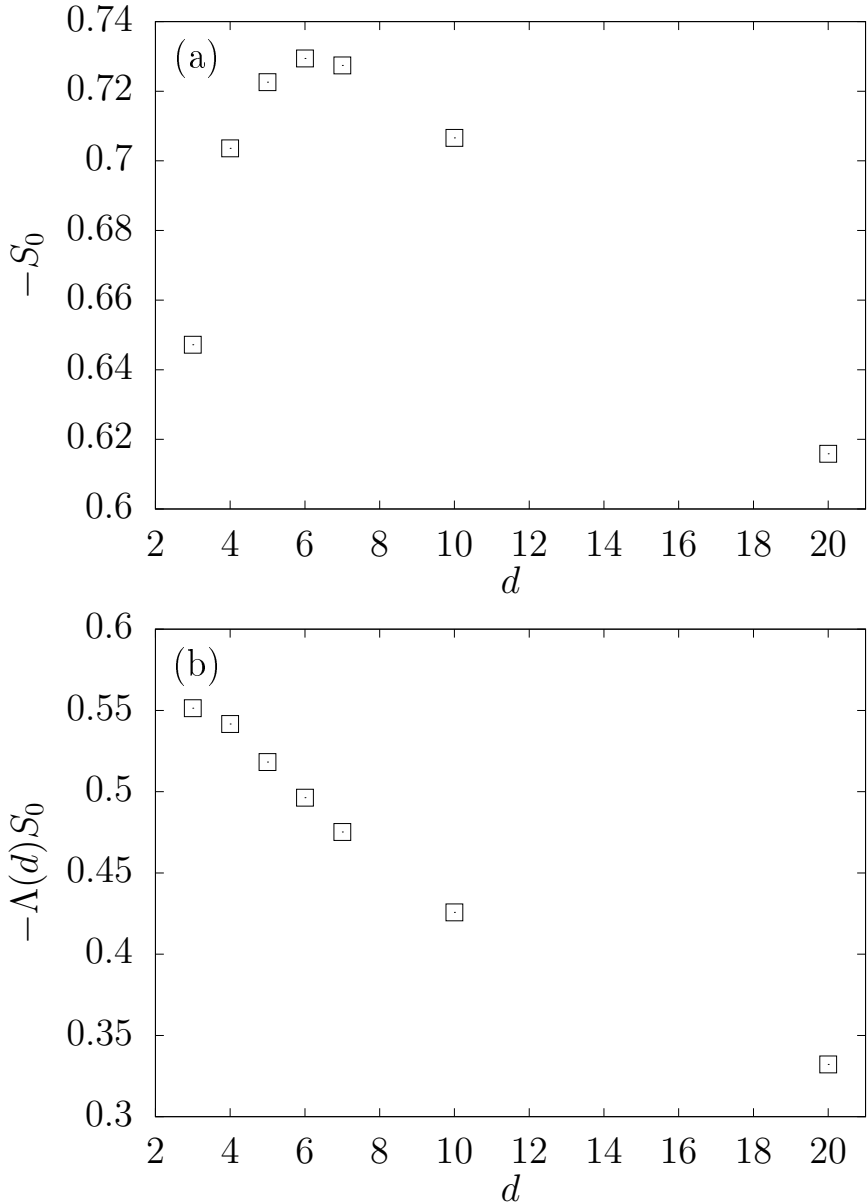}
    \caption{(a) EDQNM Velocity derivative skewness vs dimension for $C_d =1$ runs. (b) Enstrophy production term $-S_{0} \Lambda(d)$ vs dimension for $C_d = 1$ runs.}
    \label{skew1}
\end{figure}

\section{Simulation Parameters}\label{appB}

For completeness, here we present the values used for the Kolmogorov constant in each dimension alongside the corresponding value of the free parameter. For the Kolmogorov constant the values were obtained from DNS results for three and four dimensions \citep{berera2019d} whilst in higher dimensions results obtained in \citep{gotoh2007statistical} by using the Lagrangian renormalised approximation (LRA) \citep{kaneda1981renormalized} are used. The appropriate value for the free parameter in each dimension is obtained using the method in Appendix \ref{appA}. These values are displayed in table 1. Additionally we give both the integral length scale and Taylor length scale Reynolds numbers for the largest cases in each dimension.

\section{Influence of the Free Parameter on the velocity derivative skewness}\label{appC}

When using the $d$-dimensional skewness equation given in equation \ref{dskewness} to measure the velocity derivative skewness in our EDQNM simulations we should be careful in our interpretation of the results. The major concern we have is that since this equation depends on $E(k)$ then the adjustable parameter $\lambda_1$, through its influence on the Kolmogorov constant, may be the cause of any trend we observe. This is an important point, as the values for the Kolmogorov constant in higher dimensions cannot be known. Indeed, in this work the values used are obtained from the self consistent LRA closure. Under this closure approximation the Kolmogorov constant decreases from three dimensions before achieving an asymptotic constant value at high dimension. 

To understand the influence of $\lambda_!$ on our results we have performed another set of simulations where $\lambda_1$ is chosen such that $C_d = 1$ in all cases. The results of these simulations are presented in figure \ref{skew1}. In figure \ref{skew1}a it is clear that even when the influence of $\lambda_1$ on the free parameter is removed the skewness still looks to decrease at high dimension after a peak somewhere below 10 dimensions. Without data at far higher dimension we cannot speculate on the asymptotic behaviour of the velocity derivative skewness. What we can say is that, at least for the EDQNM closure, a peak value is observed which is is solely determined by the triadic interactions of the system. The significance, if there is any, of this peak skewness dimension is unknown.

\bibliography{ref.bib}

\begin{thebibliography}{76}
\expandafter\ifx\csname natexlab\endcsname\relax\def\natexlab#1{#1}\fi
\def\au#1{#1} \def\ed#1{#1} \def\yr#1{#1}\def\at#1{#1}\def\jt#1{\textit{#1}}
  \def\bt#1{#1}\def\bvol#1{\textbf{#1}} \def\vol#1{#1} \def\pg#1{#1}
  \def\publ#1{#1}\def\arxiv#1{#1}\def\org#1{#1}\def\st#1{\textit{#1}}

\bibitem[Aji \& Goldenfeld(2001)]{aji2001fluctuations}
{\sc \au{Aji, V} \& \au{Goldenfeld, N}} \yr{2001}  \at{Fluctuations in finite
  critical and turbulent systems}.  \jt{Physical Review Letters}
  \bvol{86}~(6),  \pg{1007}.

\bibitem[Andr{\'e} \& Lesieur(1977)]{andre1977influence}
{\sc \au{Andr{\'e}, JC} \& \au{Lesieur, M}} \yr{1977}  \at{Influence of
  helicity on the evolution of isotropic turbulence at high reynolds number}.
  \jt{Journal of Fluid Mechanics}  \bvol{81}~(1),  \pg{187--207}.

\bibitem[Anselmet {\em et~al.\/}(1984)Anselmet, Gagne, Hopfinger \&
  Antonia]{anselmet1984high}
{\sc \au{Anselmet, F}, \au{Gagne, YL}, \au{Hopfinger, EJ} \& \au{Antonia, RA}}
  \yr{1984}  \at{High-order velocity structure functions in turbulent shear
  flows}.  \jt{Journal of Fluid Mechanics}  \bvol{140},  \pg{63--89}.

\bibitem[Antonia \& Burattini(2006)]{antonia2006approach}
{\sc \au{Antonia, RA} \& \au{Burattini, Paolo}} \yr{2006}  \at{Approach to the
  4/5 law in homogeneous isotropic turbulence}.  \jt{Journal of fluid
  mechanics}  \bvol{550},  \pg{175}.

\bibitem[Antonia {\em et~al.\/}(2019)Antonia, Tang, Djenidi \&
  Zhou]{antonia2019finite}
{\sc \au{Antonia, RA}, \au{Tang, SL}, \au{Djenidi, L} \& \au{Zhou, Y}}
  \yr{2019}  \at{Finite reynolds number effect and the 4/5 law}.  \jt{Physical
  Review Fluids}  \bvol{4}~(8),  \pg{084602}.

\bibitem[Batchelor(1953)]{batchelor1953theory}
{\sc \au{Batchelor, GK}} \yr{1953} {\em The theory of homogeneous
  turbulence\/}.  \publ{Cambridge university press}.

\bibitem[Bell \& Nelkin(1978)]{bell1978time}
{\sc \au{Bell, TL} \& \au{Nelkin, M}} \yr{1978}  \at{Time-dependent scaling
  relations and a cascade model of turbulence}.  \jt{Journal of Fluid
  Mechanics}  \bvol{88}~(2),  \pg{369--391}.

\bibitem[Benzi {\em et~al.\/}(1984)Benzi, Paladin, Parisi \&
  Vulpiani]{benzi1984multifractal}
{\sc \au{Benzi, R}, \au{Paladin, G}, \au{Parisi, G} \& \au{Vulpiani, A}}
  \yr{1984}  \at{On the multifractal nature of fully developed turbulence and
  chaotic systems}.  \jt{Journal of Physics A: Mathematical and General}
  \bvol{17}~(18),  \pg{3521}.

\bibitem[Berera {\em et~al.\/}(2020)Berera, Ho \& Clark]{berera2019d}
{\sc \au{Berera, A}, \au{Ho, RDJG} \& \au{Clark, D}} \yr{2020}  \at{Homogeneous
  isotropic turbulence in four spatial dimensions - arxiv:2007.10953}.
  \jt{Physics of Fluids}  \bvol{32}.

\bibitem[Bos {\em et~al.\/}(2012)Bos, Chevillard, Scott \&
  Rubinstein]{bos2012reynolds}
{\sc \au{Bos, WJT}, \au{Chevillard, L}, \au{Scott, JF} \& \au{Rubinstein, R}}
  \yr{2012}  \at{Reynolds number effect on the velocity increment skewness in
  isotropic turbulence}.  \jt{Physics of Fluids}  \bvol{24}~(1),  \pg{015108}.

\bibitem[Bos \& Rubinstein(2013)]{bos2013strength}
{\sc \au{Bos, WJT} \& \au{Rubinstein, R}} \yr{2013}  \at{On the strength of the
  nonlinearity in isotropic turbulence}.  \jt{Journal of Fluid Mechanics}
  \bvol{733},  \pg{158}.

\bibitem[Bos {\em et~al.\/}(2007)Bos, Shao \& Bertoglio]{bos2007spectral}
{\sc \au{Bos, WJT}, \au{Shao, L} \& \au{Bertoglio, JP}} \yr{2007}  \at{Spectral
  imbalance and the normalized dissipation rate of turbulence}.  \jt{Physics of
  Fluids}  \bvol{19}~(4),  \pg{045101}.

\bibitem[Bowman(1996)]{bowman1996wavenumber}
{\sc \au{Bowman, JC}} \yr{1996}  \at{A wavenumber partitioning scheme for
  two-dimensional statistical closures}.  \jt{Journal of Scientific Computing}
  \bvol{11}~(4),  \pg{343--372}.

\bibitem[Bramwell {\em et~al.\/}(1998)Bramwell, Holdsworth \&
  Pinton]{bramwell1998universality}
{\sc \au{Bramwell, ST}, \au{Holdsworth, PCW} \& \au{Pinton, J-F}} \yr{1998}
  \at{Universality of rare fluctuations in turbulence and critical phenomena}.
  \jt{Nature}  \bvol{396}~(6711),  \pg{552}.

\bibitem[Burattini {\em et~al.\/}(2005)Burattini, Lavoie \&
  Antonia]{burattini2005normalized}
{\sc \au{Burattini, P}, \au{Lavoie, P} \& \au{Antonia, RA}} \yr{2005}  \at{On
  the normalized turbulent energy dissipation rate}.  \jt{Physics of Fluids}
  \bvol{17}~(9),  \pg{098103}.

\bibitem[Cerbus \& Chakraborty(2017)]{cerbus2017third}
{\sc \au{Cerbus, Rory~T} \& \au{Chakraborty, Pinaki}} \yr{2017}  \at{The
  third-order structure function in two dimensions: The rashomon effect}.
  \jt{Physics of Fluids}  \bvol{29}~(11),  \pg{111110}.

\bibitem[Chen {\em et~al.\/}(1989)Chen, Herring, Kerr \&
  Kraichnan]{chen1989non}
{\sc \au{Chen, H}, \au{Herring, JR}, \au{Kerr, RM} \& \au{Kraichnan, RH}}
  \yr{1989}  \at{Non-gaussian statistics in isotropic turbulence}.  \jt{Physics
  of Fluids A: Fluid Dynamics}  \bvol{1}~(11),  \pg{1844--1854}.

\bibitem[Clark(2019)]{clark2019edqnm}
{\sc \au{Clark, D}} \yr{2019}  \at{Edinqnm code documentation} .

\bibitem[Davidson(2015)]{davidson2015turbulence}
{\sc \au{Davidson, P}} \yr{2015} {\em Turbulence: an introduction for
  scientists and engineers\/}.  \publ{Oxford University Press}.

\bibitem[De~Gennes(1975)]{de1975phase}
{\sc \au{De~Gennes, PG}} \yr{1975}  \at{Phase transition and turbulence: An
  introduction}.  \bt{In {\em Fluctuations, instabilities, and phase
  transitions\/}},  \pg{pp. 1--18}.  \publ{Springer}.

\bibitem[Doering \& Foias(2002)]{doering2002energy}
{\sc \au{Doering, CR} \& \au{Foias, C}} \yr{2002}  \at{Energy dissipation in
  body-forced turbulence}.  \jt{Journal of Fluid Mechanics}  \bvol{467},
  \pg{289--306}.

\bibitem[Edwards(1964)]{edwards1964statistical}
{\sc \au{Edwards, SF}} \yr{1964}  \at{The statistical dynamics of homogeneous
  turbulence}.  \jt{Journal of Fluid Mechanics}  \bvol{18}~(2),  \pg{239--273}.

\bibitem[Falkovich(1994)]{falkovich1994bottleneck}
{\sc \au{Falkovich, G}} \yr{1994}  \at{Bottleneck phenomenon in developed
  turbulence}.  \jt{Physics of Fluids}  \bvol{6}~(4),  \pg{1411--1414}.

\bibitem[Fournier \& Frisch(1978)]{fournier1978d}
{\sc \au{Fournier, JD} \& \au{Frisch, U}} \yr{1978}  \at{d-dimensional
  turbulence}.  \jt{Physical Review A}  \bvol{17}~(2),  \pg{747}.

\bibitem[Fournier {\em et~al.\/}(1978)Fournier, Frisch \&
  Rose]{fournier1978infinite}
{\sc \au{Fournier, JD}, \au{Frisch, U} \& \au{Rose, HA}} \yr{1978}
  \at{Infinite-dimensional turbulence}.  \jt{Journal of Physics A: Mathematical
  and General}  \bvol{11}~(1),  \pg{187}.

\bibitem[Frisch {\em et~al.\/}(2012)Frisch, Pomyalov, Procaccia \&
  Ray]{frisch2012turbulence}
{\sc \au{Frisch, U}, \au{Pomyalov, A}, \au{Procaccia, I} \& \au{Ray, SS}}
  \yr{2012}  \at{Turbulence in noninteger dimensions by fractal fourier
  decimation}.  \jt{Physical review letters}  \bvol{108}~(7),  \pg{074501}.

\bibitem[Frisch {\em et~al.\/}(1978)Frisch, Sulem \& Nelkin]{frisch1978simple}
{\sc \au{Frisch, U}, \au{Sulem, P-L} \& \au{Nelkin, M}} \yr{1978}  \at{A simple
  dynamical model of intermittent fully developed turbulence}.  \jt{Journal of
  Fluid Mechanics}  \bvol{87}~(4),  \pg{719--736}.

\bibitem[Ginzburg(1960)]{ginzburg1960vl}
{\sc \au{Ginzburg, VL}} \at{ \yr{1960} } \jt{Fiz. Tverd. Tela}  \bvol{2},
  \pg{2031}.

\bibitem[Giuliani {\em et~al.\/}(2002)Giuliani, Jensen \&
  Yakhot]{giuliani2002critical}
{\sc \au{Giuliani, P}, \au{Jensen, MH} \& \au{Yakhot, V}} \yr{2002}
  \at{Critical “dimension” in shell model turbulence}.  \jt{Physical Review
  E}  \bvol{65}~(3),  \pg{036305}.

\bibitem[Gotoh {\em et~al.\/}(2002)Gotoh, Fukayama \&
  Nakano]{gotoh2002velocity}
{\sc \au{Gotoh, T}, \au{Fukayama, D} \& \au{Nakano, T}} \yr{2002}  \at{Velocity
  field statistics in homogeneous steady turbulence obtained using a
  high-resolution direct numerical simulation}.  \jt{Physics of Fluids}
  \bvol{14}~(3),  \pg{1065--1081}.

\bibitem[Gotoh {\em et~al.\/}(2007)Gotoh, Watanabe, Shiga, Nakano \&
  Suzuki]{gotoh2007statistical}
{\sc \au{Gotoh, T}, \au{Watanabe, Y}, \au{Shiga, Y}, \au{Nakano, T} \&
  \au{Suzuki, E}} \yr{2007}  \at{Statistical properties of four-dimensional
  turbulence}.  \jt{Physical Review E}  \bvol{75}~(1),  \pg{016310}.

\bibitem[Herring {\em et~al.\/}(1982)Herring, Schertzer, Lesieur, Newman,
  Chollet \& Larcheveque]{herring1982comparative}
{\sc \au{Herring, JR}, \au{Schertzer, D}, \au{Lesieur, M}, \au{Newman, GR},
  \au{Chollet, JP} \& \au{Larcheveque, M}} \yr{1982}  \at{A comparative
  assessment of spectral closures as applied to passive scalar diffusion}.
  \jt{Journal of Fluid Mechanics}  \bvol{124},  \pg{411--437}.

\bibitem[Ishihara {\em et~al.\/}(2009)Ishihara, Gotoh \&
  Kaneda]{ishihara2009study}
{\sc \au{Ishihara, T}, \au{Gotoh, T} \& \au{Kaneda, Y}} \yr{2009}  \at{Study of
  high--reynolds number isotropic turbulence by direct numerical simulation}.
  \jt{Annual Review of Fluid Mechanics}  \bvol{41},  \pg{165--180}.

\bibitem[Ishihara {\em et~al.\/}(2016)Ishihara, Morishita, Yokokawa, Uno \&
  Kaneda]{ishihara2016energy}
{\sc \au{Ishihara, T}, \au{Morishita, K}, \au{Yokokawa, M}, \au{Uno, A} \&
  \au{Kaneda, Y}} \yr{2016}  \at{Energy spectrum in high-resolution direct
  numerical simulations of turbulence}.  \jt{Physical Review Fluids}
  \bvol{1}~(8),  \pg{082403}.

\bibitem[Kaneda(1981)]{kaneda1981renormalized}
{\sc \au{Kaneda, Y}} \yr{1981}  \at{Renormalized expansions in the theory of
  turbulence with the use of the lagrangian position function}.  \jt{Journal of
  Fluid Mechanics}  \bvol{107},  \pg{131--145}.

\bibitem[Kaneda {\em et~al.\/}(2003)Kaneda, Ishihara, Yokokawa, Itakura \&
  Uno]{kaneda2003energy}
{\sc \au{Kaneda, Y}, \au{Ishihara, T}, \au{Yokokawa, M}, \au{Itakura, K} \&
  \au{Uno, A}} \yr{2003}  \at{Energy dissipation rate and energy spectrum in
  high resolution direct numerical simulations of turbulence in a periodic
  box}.  \jt{Physics of Fluids}  \bvol{15}~(2),  \pg{L21--L24}.

\bibitem[von K{\'a}rm{\'a}n \& Howarth(1938)]{de1938statistical}
{\sc \au{von K{\'a}rm{\'a}n, T} \& \au{Howarth, L}} \yr{1938}  \at{On the
  statistical theory of isotropic turbulence}.  \jt{Proceedings of the Royal
  Society of London. Series A-Mathematical and Physical Sciences}
  \bvol{164}~(917),  \pg{192--215}.

\bibitem[Kerr(1990)]{kerr1990velocity}
{\sc \au{Kerr, RM}} \yr{1990}  \at{Velocity, scalar and transfer spectra in
  numerical turbulence}.  \jt{Journal of Fluid Mechanics}  \bvol{211},
  \pg{309--332}.

\bibitem[Kolmogorov(1941{\natexlab{{\em a\/}}})]{kolmogorov1941dissipation}
{\sc \au{Kolmogorov, AN}} \yr{1941{\natexlab{{\em a\/}}}} Dissipation of energy
  in locally isotropic turbulence.  \bt{In {\em Akademiia Nauk SSSR
  Doklady\/}}, ,  \vol{vol.~32},  \pg{p.~16}.

\bibitem[Kolmogorov(1941{\natexlab{{\em b\/}}})]{kolmogorov1941local}
{\sc \au{Kolmogorov, AN}} \yr{1941{\natexlab{{\em b\/}}}} The local structure
  of turbulence in incompressible viscous fluid for very large reynolds'
  numbers.  \bt{In {\em Akademiia Nauk SSSR Doklady\/}}, ,  \vol{vol.~30},
  \pg{pp. 301--305}.

\bibitem[Kolmogorov(1941{\natexlab{{\em c\/}}})]{kolmogorov1941degeneration}
{\sc \au{Kolmogorov, AN}} \yr{1941{\natexlab{{\em c\/}}}} On the degeneration
  of isotropic turbulence in an incompressible viscous fluid.  \bt{In {\em
  Dokl. Akad. Nauk SSSR\/}}, ,  \vol{vol.~31},  \pg{pp. 319--323}.

\bibitem[Kolmogorov(1962)]{kolmogorov1962refinement}
{\sc \au{Kolmogorov, AN}} \yr{1962}  \at{A refinement of previous hypotheses
  concerning the local structure of turbulence in a viscous incompressible
  fluid at high reynolds number}.  \jt{Journal of Fluid Mechanics}
  \bvol{13}~(1),  \pg{82--85}.

\bibitem[Kraichnan(1959)]{kraichnan1959structure}
{\sc \au{Kraichnan, RH}} \yr{1959}  \at{The structure of isotropic turbulence
  at very high reynolds numbers}.  \jt{Journal of Fluid Mechanics}
  \bvol{5}~(4),  \pg{497--543}.

\bibitem[Kraichnan(1974)]{kraichnan1974convection}
{\sc \au{Kraichnan, RH}} \yr{1974}  \at{Convection of a passive scalar by a
  quasi-uniform random straining field}.  \jt{Journal of Fluid Mechanics}
  \bvol{64}~(4),  \pg{737--762}.

\bibitem[Leith(1971)]{leith1971atmospheric}
{\sc \au{Leith, CE}} \yr{1971}  \at{Atmospheric predictability and
  two-dimensional turbulence}.  \jt{Journal of the Atmospheric Sciences}
  \bvol{28}~(2),  \pg{145--161}.

\bibitem[Leith \& Kraichnan(1972)]{leith1972predictability}
{\sc \au{Leith, CE} \& \au{Kraichnan, RH}} \yr{1972}  \at{Predictability of
  turbulent flows}.  \jt{Journal of the Atmospheric Sciences}  \bvol{29}~(6),
  \pg{1041--1058}.

\bibitem[Lesieur(1987)]{lesieur1987turbulence}
{\sc \au{Lesieur, M}} \yr{1987} {\em Turbulence in fluids: stochastic and
  numerical modelling\/}.  \publ{Nijhoff Boston, MA}.

\bibitem[Lesieur \& Schertzer(1978)]{lesieur1978amortissement}
{\sc \au{Lesieur, M} \& \au{Schertzer, D}} \yr{1978}  \at{Self-similar damping
  of a large reynolds number turbulence}.  \jt{J. de Mecanique,}
  \bvol{17}~(4),  \pg{609--646}.

\bibitem[Liao(1990)]{liao1990some}
{\sc \au{Liao, W}} \yr{1990}  \at{Some ideas on the freely decaying
  navier-stokes turbulence}.  \jt{Journal of Physics A: Mathematical and
  General}  \bvol{23}~(4),  \pg{L159}.

\bibitem[Liao(1991)]{liao1991kolmogorov}
{\sc \au{Liao, W}} \yr{1991}  \at{Kolmogorov exponents for near-incompressible
  turbulence from perturbative quantum field theory}.  \jt{Journal of
  statistical physics}  \bvol{65}~(1-2),  \pg{1--32}.

\bibitem[McComb(1990)]{mccomb1990physics}
{\sc \au{McComb, WD}} \yr{1990}  \at{The physics of fluid turbulence}.
  \jt{Chemical physics} .

\bibitem[McComb(2014)]{mccomb2014homogeneous}
{\sc \au{McComb, WD}} \yr{2014} {\em Homogeneous, isotropic turbulence:
  phenomenology, renormalization and statistical closures\/}, ,  \vol{vol.
  162}.  \publ{OUP Oxford}.

\bibitem[McComb {\em et~al.\/}(2015)McComb, Berera, Yoffe \&
  Linkmann]{mccomb2015energy}
{\sc \au{McComb, WD}, \au{Berera, A}, \au{Yoffe, SR} \& \au{Linkmann, MF}}
  \yr{2015}  \at{Energy transfer and dissipation in forced isotropic
  turbulence}.  \jt{Physical Review E}  \bvol{91}~(4),  \pg{043013}.

\bibitem[Mestayer(1982)]{mestayer1982local}
{\sc \au{Mestayer, P}} \yr{1982}  \at{Local isotropy and anisotropy in a
  high-reynolds-number turbulent boundary layer}.  \jt{Journal of Fluid
  Mechanics}  \bvol{125},  \pg{475--503}.

\bibitem[Millionshchikov(1941)]{millionshchikov1941theory}
{\sc \au{Millionshchikov, MD}} \yr{1941} On the theory of homogeneous isotropic
  turbulence.  \bt{In {\em Dokl. Akad. Nauk SSSR\/}}, ,  \vol{vol.~32},
  \pg{pp. 611--614}.

\bibitem[Nelkin(1974)]{nelkin1974turbulence}
{\sc \au{Nelkin, M}} \yr{1974}  \at{Turbulence, critical fluctuations, and
  intermittency}.  \jt{Physical Review A}  \bvol{9}~(1),  \pg{388}.

\bibitem[Nelkin(2001)]{nelkin2001does}
{\sc \au{Nelkin, M}} \yr{2001}  \at{Does kolmogorov mean field theory become
  exact for turbulence above some critical dimension?}  \jt{arXiv preprint
  nlin/0103046} .

\bibitem[Orszag(1970)]{orszag1970analytical}
{\sc \au{Orszag, SA}} \yr{1970}  \at{Analytical theories of turbulence}.
  \jt{Journal of Fluid Mechanics}  \bvol{41}~(2),  \pg{363--386}.

\bibitem[Orszag(1974)]{orszag1974lectures}
{\sc \au{Orszag, SA}} \yr{1974} {\em Lectures on the statistical theory of
  turbulence\/}, ,  \vol{vol.~7}.  \publ{Flow Research Incorporated}.

\bibitem[Qian(1997)]{qian1997inertial}
{\sc \au{Qian, J}} \yr{1997}  \at{Inertial range and the finite reynolds number
  effect of turbulence}.  \jt{Physical Review E}  \bvol{55}~(1),  \pg{337}.

\bibitem[Qian(1999)]{qian1999slow}
{\sc \au{Qian, J}} \yr{1999}  \at{Slow decay of the finite reynolds number
  effect of turbulence}.  \jt{Physical Review E}  \bvol{60}~(3),  \pg{3409}.

\bibitem[Rose \& Sulem(1978)]{rose1978fully}
{\sc \au{Rose, HA} \& \au{Sulem, P-L}} \yr{1978}  \at{Fully developed
  turbulence and statistical mechanics}.  \jt{Journal de Physique}
  \bvol{39}~(5),  \pg{441--484}.

\bibitem[Saddoughi \& Veeravalli(1994)]{saddoughi1994local}
{\sc \au{Saddoughi, SG} \& \au{Veeravalli, SV}} \yr{1994}  \at{Local isotropy
  in turbulent boundary layers at high reynolds number}.  \jt{Journal of Fluid
  Mechanics}  \bvol{268},  \pg{333--372}.

\bibitem[Sagaut \& Cambon(2008)]{sagaut2008homogeneous}
{\sc \au{Sagaut, Pierre} \& \au{Cambon, Claude}} \yr{2008} {\em Homogeneous
  turbulence dynamics\/}, ,  \vol{vol.~10}.  \publ{Springer}.

\bibitem[Siggia(1977)]{siggia1977origin}
{\sc \au{Siggia, ED}} \yr{1977}  \at{Origin of intermittency in fully developed
  turbulence}.  \jt{Physical Review A}  \bvol{15}~(4),  \pg{1730}.

\bibitem[Sreenivasan(1984)]{sreenivasan1984scaling}
{\sc \au{Sreenivasan, KR}} \yr{1984}  \at{On the scaling of the turbulence
  energy dissipation rate}.  \jt{The Physics of fluids}  \bvol{27}~(5),
  \pg{1048--1051}.

\bibitem[Suzuki {\em et~al.\/}(2005)Suzuki, Nakano, Takahashi \&
  Gotoh]{suzuki2005energy}
{\sc \au{Suzuki, E}, \au{Nakano, T}, \au{Takahashi, N} \& \au{Gotoh, T}}
  \yr{2005}  \at{Energy transfer and intermittency in four-dimensional
  turbulence}.  \jt{Physics of Fluids}  \bvol{17}~(8),  \pg{081702}.

\bibitem[Tang {\em et~al.\/}(2017)Tang, Antonia, Djenidi, Danaila \&
  Zhou]{tang2017finite}
{\sc \au{Tang, SL}, \au{Antonia, RA}, \au{Djenidi, L}, \au{Danaila, L} \&
  \au{Zhou, Y}} \yr{2017}  \at{Finite reynolds number effect on the scaling
  range behaviour of turbulent longitudinal velocity structure functions}.
  \jt{Journal of Fluid Mechanics}  \bvol{820},  \pg{341}.

\bibitem[Taylor(1935)]{taylor1935statistical}
{\sc \au{Taylor, GI}} \yr{1935}  \at{Statistical theory of turbulence}.
  \jt{Proceedings of the Royal Society of London. Series A-Mathematical and
  Physical Sciences}  \bvol{151}~(873),  \pg{421--444}.

\bibitem[Tchoufag {\em et~al.\/}(2012)Tchoufag, Sagaut \&
  Cambon]{tchoufag2012spectral}
{\sc \au{Tchoufag, J}, \au{Sagaut, Pierre} \& \au{Cambon, Claude}} \yr{2012}
  \at{Spectral approach to finite reynolds number effects on kolmogorov’s 4/5
  law in isotropic turbulence}.  \jt{Physics of Fluids}  \bvol{24}~(1),
  \pg{015107}.

\bibitem[Wang {\em et~al.\/}(1996)Wang, Chen, Brasseur \&
  Wyngaard]{wang1996examination}
{\sc \au{Wang, LP}, \au{Chen, S}, \au{Brasseur, JG} \& \au{Wyngaard, JC}}
  \yr{1996}  \at{Examination of hypotheses in the kolmogorov refined turbulence
  theory through high-resolution simulations. part 1. velocity field}.
  \jt{Journal of Fluid Mechanics}  \bvol{309},  \pg{113--156}.

\bibitem[Wilson(1971)]{wilson1971renormalization}
{\sc \au{Wilson, KG}} \yr{1971}  \at{Renormalization group and critical
  phenomena. i. renormalization group and the kadanoff scaling picture}.
  \jt{Physical review B}  \bvol{4}~(9),  \pg{3174}.

\bibitem[Wilson \& Fisher(1972)]{wilson1972critical}
{\sc \au{Wilson, KG} \& \au{Fisher, ME}} \yr{1972}  \at{Critical exponents in
  3.99 dimensions}.  \jt{Physical Review Letters}  \bvol{28}~(4),  \pg{240}.

\bibitem[Wyld~Jr(1961)]{wyld1961formulation}
{\sc \au{Wyld~Jr, HW}} \yr{1961}  \at{Formulation of the theory of turbulence
  in an incompressible fluid}.  \jt{Annals of Physics}  \bvol{14},
  \pg{143--165}.

\bibitem[Yakhot(2001)]{yakhot2001mean}
{\sc \au{Yakhot, V}} \yr{2001}  \at{Mean-field approximation and a small
  parameter in turbulence theory}.  \jt{Physical Review E}  \bvol{63}~(2),
  \pg{026307}.

\bibitem[Yamamoto {\em et~al.\/}(2012)Yamamoto, Shimizu, Inoshita, Nakano \&
  Gotoh]{yamamoto2012local}
{\sc \au{Yamamoto, T}, \au{Shimizu, H}, \au{Inoshita, T}, \au{Nakano, T} \&
  \au{Gotoh, T}} \yr{2012}  \at{Local flow structure of turbulence in three,
  four, and five dimensions}.  \jt{Physical Review E}  \bvol{86}~(4),
  \pg{046320}.

\end{thebibliography}
\bibliographystyle{jfm}

\end{document}